\documentclass{aa}  
\usepackage{graphicx}
\usepackage{txfonts}
\usepackage[utf8]{inputenc}
\usepackage{booktabs}
\usepackage{multirow}
\usepackage{siunitx}
\usepackage{amsmath}
\usepackage{setspace}
\usepackage{makecell}

\usepackage[dvipsnames]{xcolor}

\begin{document}

   \title{How a Close-in Planet Protects its White Dwarf Host from Pollution}


   \author{
  Xin-Yue Zhang\inst{1,2}
  \and
  Ji-Wei Xie\inst{1,2}
  \and
  Di-Chang Chen\inst{3}
  \and
  Ji-Lin Zhou\inst{1,2}
}
   \institute{School of Astronomy and Space Science, Nanjing University, Nanjing 210023, People’s Republic of China\\
             \email{jwxie@nju.edu.cn}
             \and
             Key Laboratory of Modern Astronomy and Astrophysics, Ministry of Education, Nanjing 210023, PR China
             \and
  School of Physics and Astronomy, Sun Yat-sen University, DaXue Road 2, 519082, Zhuhai, China }

   \date{Received xxx xx, xxxx; accepted xxx xx, xxxx}

 
  \abstract
   {Approximately 25–50\% of white dwarfs (WDs) exhibit metal absorption lines in their photospheres, which are attributed to accretion from their remnant planetary systems. Although white dwarfs with detected planetary systems are more likely to show photospheric pollution, one notable exception—WD 1856+534—host close-in giant planet yet exhibit no detectable photospheric metal pollution. Previous studies have proposed that massive, close-in planets can block inward transport of small particles driven by radiative forces (e.g., Poynting–Robertson drag and the Yarkovsky effect). However, it remains unclear whether the close-in planet can similarly prevent delivery of larger bodies via dynamical interactions. }
   { We aim to quantify the protective influence of close-in planets on white-dwarf pollution by asteroids approaching on near-parabolic orbits, and to explore the planetary masses and orbital separations required to provide effective protection. }
   {We perform ensembles of short-term N-body integrations, sampling a range of planet masses and orbital separations and initializing asteroids on highly eccentric orbits with periapses near the WD Roche radius, in order to measure scattering, capture, and ejection outcomes and quantify the planet’s shielding efficiency.}
   {For WD1856+534b-like configurations ($a_p=0.02$ au), giant planets with masses greater than 0.5 Jupiter masses are sufficient to clear over 80\% of highly eccentric small-body contaminants. The effectiveness of the protective effect diminishes with decreasing planetary mass and increasing semi-major axis. These findings help explain why some white dwarfs that host close-in giant planets do not show the photospheric metal pollution commonly observed in other systems.}
   {}

   \keywords{white dwarfs — planetary systems — methods: numerical — planets and satellites: dynamical evolution and stability
               }

   \maketitle
%

\section{Introduction}
To date, over 6,000 exoplanets have been discovered around main-sequence stars. On average, most solar-type stars in the Milky Way host at least one planet \citep{zhu_exoplanet_2021}. 
As most known host stars of a planetary system will end their life as white dwarfs (WDs), these post-main sequence stars offer a unique window to understand the composition and evolution of exoplanetary systems \citep{mcdonald_binary_2023}. 

Owing to the large surface gravities of white dwarfs, heavy elements (metals) undergo rapid diffusive settling in their photospheres on timescales that are short compared with typical WD cooling ages \citep{paquette_diffusion_1986}. Consequently, metals accreted from external reservoirs are readily detectable against otherwise pure hydrogen- or helium-dominated WD atmospheres; this phenomenon is referred to as ``white dwarf pollution" \citep{veras_planetary_2021}. Spectroscopic surveys show that approximately 25\%–50\% of surveyed white dwarfs display signatures of metal pollution \citep{zuckerman_metal_2003,zuckerman_ancient_2010,koester_frequency_2014,coutu_analysis_2019},  implying that a substantial fraction of WDs are presently accreting external material. 

Composition analyses of polluted white dwarfs indicate that the abundances of accreted elements are broadly consistent with the bulk compositions of Solar System asteroids and comets \citep[e.g.,]{jura_carbon_2006,jura_x-ray_2009,zuckerman_chemical_2007,zuckerman_aluminumcalcium-rich_2011,swan_interpretation_2019,swan_planetesimals_2023,xu__chemical_2017}. These planetesimal-derived pollutants are driven within the white dwarf’s Roche sphere, where they undergo tidal disruption and their fragments are subsequently accreted onto the stellar surface \citep[e.g.,]{jura_tidally_2003,debes_link_2012,chen_power-law_2019}. The delivery mechanisms that bring material to this state depend strongly on size: small particles (size $\sim 10^{-6}-10^{3}m$) are transported inward primarily by radiative forces such as Poynting–Robertson (PR) drag and the Yarkovsky effect \citep{veras_orbit_2022}, whereas large bodies ($\sim 10^{3}-10^6m$) must be driven to very high eccentricities—typically through planetary perturbations or interactions with stellar/companion bodies—in order to approach the Roche radius \citep[e.g.,]{debes_are_2002,frewen_eccentric_2014,mcdonald_binary_2023,veras_smallest_2023}.
 
Planets orbiting white dwarfs are widely invoked as dynamical agents that can deliver small bodies into star-grazing orbits and thereby contribute to photospheric pollution \citep[e.g.,]{frewen_eccentric_2014,veras_smallest_2023}. Nevertheless, only a handful of \textcolor{black}{planetary mass objects} have been confirmed in post–main-sequence systems to date (e.g., PSR B1620-26(AB)b: \cite{thorsett_psr_1993,sigurdsson_young_2003}, WD 0806-661b: \cite{luhman_discovery_2011}, WD J0914+1914b: \cite{gansicke_accretion_2019}, WD 1856+534b: \cite{vanderburg_giant_2020}, \textcolor{black}{MOA-2010-BLG-477Lb: \cite{blackman_jovian_2021}, PHL 5038AB: \cite{casewell_phl_2024})}, with several additional candidate detections recently reported from JWST \citep{mullally_jwst_2024,limbach_miri_2024,debes_metal_2025}. While white dwarfs with planetary systems are generally more susceptible to pollution, \textcolor{black}{the well-studied system— WD 1856+534 (hosting a close-in giant at $\approx 0.02$ au)—show no detectable photospheric metal pollution despite the presence of massive, short-period companions \citep{vanderburg_giant_2020,xu_geminigmos_2021}.} These observations motivate the hypothesis that sufficiently massive close-in planets act as a dynamical barrier—either by ejecting, capturing, or redirecting incoming bodies—thereby suppressing the delivery of pollutant material to the white dwarf photosphere.

Previous studies have examined multiple dynamical mechanisms by which planetary companions can reduce white-dwarf photospheric pollution originating from diverse sources. For meter- to kilometer-scale particles, \citet{veras_white_2020} showed that close-in giant planets can impede inward migration of radiatively driven material—such as that caused by Poynting–Robertson drag and the Yarkovsky effect—by trapping particles into mean-motion resonances. Regarding cometary contributions, \citet{oconnor_pollution_2023} and \citet{pham_polluting_2024} estimated that planets on wide orbits ($\sim 10$–$100$ au) can reduce cometary delivery rates through a combination of direct scattering and companion-induced secular precession. Despite these advances, an important gap persists concerning the planet’s suppressive effect on the dominant pollution channel: dynamically excited, high-eccentricity asteroids that ultimately intersect the white dwarf’s Roche radius. \textcolor{black}{While specific case studies (e.g., \citealt{rogers_wd0141-675_2023}) have shown that individual planets can eject infalling bodies, a systematic quantification of whether such gravitational interactions statistically reduce the overall pollution rate across a broad planetary parameter space remains absent.}

\begin{table}[t!]
\centering
\caption{Parameters for the benchmark case in this work. We choose the upper limit of planet mass to get the strongest protection. \(T_{\mathrm{ast}}\) is the orbital period decided by the initial orbital parameters of asteroids. }\label{tabel:parameter}
\renewcommand{\arraystretch}{1.20} 
\begin{tabular}{@{} l c @{}}
\toprule
\toprule
\textbf{Stellar Mass \(\left(M_{\mathrm{WD}}\right)\)} & \(0.518\ M_{\odot}\) \\
\textbf{Planet Mass \(\left(M_{p}\right)\)}            & \(13.8\ M_{\mathrm{Jup}}\) \\
\textbf{Planet Radius \(\left(R_{p}\right)\)}            & \(10.4\ R_\oplus\) \\
\textbf{Planet semi-major axis \(\left(a_{p}\right)\)} & \(0.02\ \mathrm{au}\) \\
\textbf{Asteroid semi-major axis \(\left(a_{\mathrm{ast}}\right)\)} & \([1,10]\ \mathrm{au}\) \\
\textbf{Asteroid pericenter \(\left(q_{\mathrm{ast}}\right)\)} & \([0.5,1]\ r_{\mathrm{Roche}}\) \\
\textbf{Simulation time \((T)\)} & \(10T_{\mathrm{ast}}\) \\
\bottomrule
\bottomrule
\end{tabular}
\renewcommand{\arraystretch}{1.0}
\end{table}

In this study, we perform short-term three-body simulations to quantify how a close-in planet influences near-parabolic asteroids\footnote{Here, we use asteroids to refer to any small bodies that could pollute the white dwarf.}. We then statistically evaluate the efficiency with which various planetary companions inhibit the injection of pollutant material into the white dwarf—i.e., their effectiveness as dynamical ``protector" of the host star. In section \ref{sec:methods} we describe our physical model and numerical setup, specify the ranges of planetary and asteroid parameters explored, and define the quantitative criteria used to measure planetary ``protection". Section \ref{sec:results} presents representative case studies and a statistical analysis of population outcomes across different planetary masses and orbital separations. In section \ref{sec:discuss} we examine several factors that influence our results, discuss limitations and caveats, and compare our findings with previous studies. Finally, we summarize our main conclusions and implications in section \ref{sec:conclu}.
\section{Methods}\label{sec:methods}

\subsection{Simulation set up}\label{subsec:setup}

\begin{figure}[ht!]
\includegraphics[width=0.48\textwidth]{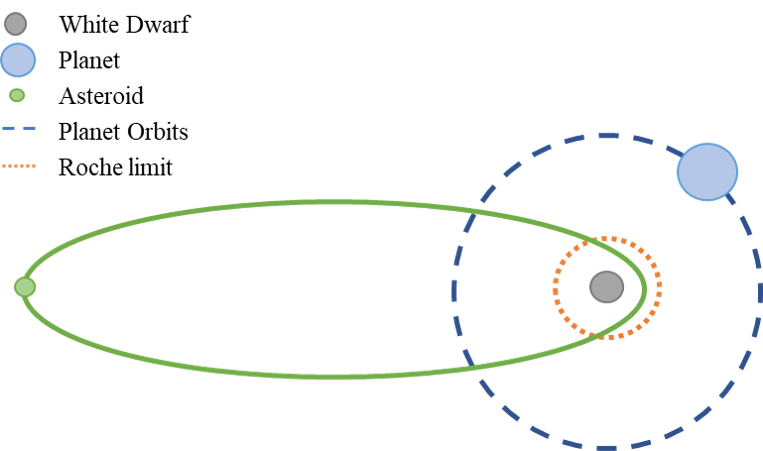}
\caption{Simplified model of the white dwarf system used in our simulation. The planet orbits the central white dwarf on a circular path, while an asteroid initially follows a highly eccentric trajectory toward the white dwarf. Please note that the figure is not to scale. \label{fig:model}}
\end{figure}

We consider a system in which a planet moves on a circular orbit around a central white dwarf while an asteroid approaches the star on a highly eccentric trajectory. Because the mass of the asteroid is negligible relative to the planet, we treat the asteroid as a massless test particle throughout the simulations. The schematic model is shown in Fig.~\ref{fig:model}. 

We set the central white dwarf mass to $M_\mathrm{WD}=0.518M_{\odot}$, adopting the best-fit mass of WD~1856+534  reported by \citet{vanderburg_giant_2020}. \textcolor{black}{We also adopted the corresponding white dwarf radius of $0.0131 R_{\odot}$ to identify asteroids that collided with the white dwarf. }Although multiple studies report marginally different stellar parameters for this system \citep{xu_geminigmos_2021,alonso_transmission_2021,limbach_thermal_2025}, tests we perform indicate that variations within the quoted uncertainty in $M_\mathrm{WD}$ do not produce statistically significant changes in our results (see section \ref{subsec:influence} for detailed statistics). 

The mass of WD1856+534b remains observationally unconstrained; theoretical models of brown-dwarf evolution and atmospheres provide an upper limit of $13.8~M_{\mathrm{Jup}}$ (95\% confidence) \citep{vanderburg_giant_2020}. More recent JWST observations \citep{limbach_thermal_2025} constrain the upper limit to approximately $6~M_{\mathrm{Jup}}$. For the primary parameter study we adopt the more conservative upper limit $13.8~M_{\mathrm{Jup}}$ in order to quantify the maximum potential protective effect of a planet; we then explore a broad range of planetary masses from earth mass to 14 Jupiter mass to assess how mass influences on our results. We set the planet’s semi-major axis ($a_p$) to 0.02 au for the benchmark case and additionally selected several values of $a_p$ between \textcolor{black}{0.02~au and 1~au} to study how orbital separation affects our outcomes. The results of the parameter study appear in section~\ref{subsec:planet}.

\textcolor{black}{Given that a giant planet's physical radius is significantly larger than that of a white dwarf, asteroid-planet collisions represent a critical sink for potential polluters. We define a collision event whenever the distance between a particle and a massive body falls below its physical radius. For our benchmark case, we adopt the planet radius $R_p=10.4~R_\oplus$ based on the observed radius of WD 1856+534b \citep{vanderburg_giant_2020}. For other planets across our explored mass range ($1~M_\oplus$ to $14~M_\mathrm{Jup}$), we estimate the physical radius using the empirical mass-radius relations from \citet{muller_mass-radius_2024}: 
\begin{equation}\label{equa:radius}
R_p =
\begin{cases}
(1.02 \pm 0.03)\,M_p^{(0.27 \pm 0.04)}, & M_p < (4.37 \pm 0.72),\\[6pt]
(0.56 \pm 0.03)\,M_p^{(0.67 \pm 0.05)}, & (4.37 \pm 0.72) < M_p < (127 \pm 7),\\[6pt]
(18.6 \pm 6.7)\,M_p^{(-0.06 \pm 0.07)}, & M_p > (127 \pm 7).
\end{cases}
\end{equation},
where $R_p$ and $M_p$ are expressed in Earth units ($R_\oplus$ and $M_\oplus$). }

We sample asteroid semi-major axes $a_\mathrm{ast}$ uniformly within [1, 10] au. \textcolor{black}{This range is chosen to represent the inner-system reservoir where the majority of rocky polluters are expected to originate, while ensuring $a_\mathrm{ast} > a_p$ for all planetary configurations ($a_p \leq 1~\mathrm{au}$), thereby guaranteeing that the asteroids’ initial orbits can cross that of the planet through subsequent eccentricity excitation. We note that the dynamical influence of the asteroid's initial semi-major axis depends on the planet's properties. For massive, close-in planets, this 1–10 au range typically sits above the threshold $a_\mathrm{in}$ (see section~\ref{subsec:influence}). However, for lower-mass or more distant planets, $a_\mathrm{in}$ may shift toward larger values, meaning their shielding effectiveness could be more sensitive to the specific spatial distribution of the asteroid belt. By adopting a fixed 1–10 au range, we provide a consistent benchmark to compare the protective capacity across the entire planetary parameter space, while acknowledging that the exact protection fractions for the least massive planets may vary if more distant reservoirs are considered. More realistic asteroid semi-major axis distributions can be set by specific pre‑excitation mechanisms, which are outside the scope of this work.}

We assume that each asteroid initially possesses a perihelion within the white dwarf's Roche limit. Specifically, we sample the initial pericenter $q_\mathrm{ast}$ uniformly in the range (0.5, 1.0) $r_\mathrm{roche}$. \textcolor{black}{This ensures that the asteroids are on near-disruptive trajectories while providing sufficient variation to sample planet-induced gravitational deflections. }The Roche radius is defined as
\begin{equation}
r_{\mathrm{Roche}}=\left(\frac{3M_{\mathrm{WD}}}{2\pi\rho_{\mathrm{ast}}}\right)^{1/3}
\end{equation},
where we adopt an average asteroid density of $\rho_{\mathrm{ast}}=2 \ g/cm^3$ \citep{carry_density_2012}, yielding $r_\mathrm{Roche} \approx 0.004$ au. \textcolor{black}{To further explore the sensitivity of our results, we also consider a population of asteroids perturbed onto orbits just exterior to the Roche limit \citep{li_can_2025} by conducting additional tests with $q_\mathrm{ast}$ sampled in (1, 2) $r_{\mathrm{Roche}}$.}

In each simulation, the asteroid is initialized at its apoastron (true anomaly $f=\pi$). The argument of pericenter $\omega$ is drawn uniformly at random from \([0,2\pi)\). Inclination \(i\) and longitude of the ascending node \(\Omega\) are sampled in two separate suites to explore both coplanar and non\text{-}coplanar configurations: in the coplanar suite we set \(i=\Omega=0\); in the non\text{-}coplanar suite we draw \(i\) and \(\Omega\) randomly from $[0,\pi]$ and $[0,2\pi)$, respectively. The dynamical differences between these configurations are discussed in section \ref{subsec:population}.

We set the simulation time in units of the asteroid's initial orbital period $T_\mathrm{ast}$, determined by the initial semi-major axis $a_\mathrm{ast}$ of the asteroid. \textcolor{black}{An ideal mechanism for a planet to prevent pollution is ejecting the asteroid before it is completely disrupted. Crucially, an asteroid entering the Roche limit on a highly eccentric orbit is expected to undergo significant tidal fragmentation and complete disintegration within approximately 10 periapsis passages \citep{malamud_tidal_2020}. Since we treat the asteroid as a single intact body to track its orbital evolution, we limit our primary simulations to $10~T_\mathrm{ast}$. For the majority of the cases studied, most ejections and collisions occur within these first few orbits. While some planets may eventually eject asteroids over much longer timescales (as discussed in section~\ref{subsec:population}),  bodies that remain bound and within the disruption zone after $10~T_\mathrm{ast}$ are assumed to be totally disrupted. Beyond this point, further orbital evolution would be dominated by the dynamics of debris streams and non-gravitational effects such as Poynting-Robertson drag \citep[e.g.,][]{li_accretion_2021}, which are beyond the scope of our current N-body model.}

Our simulations focus primarily on resolving the dynamics during periapsis. Although the Roche region occupies only a small fraction of the total orbit, the highly eccentric nature of the trajectories demands exceptional temporal resolution. This necessitates substantial computational resources, as extremely small time steps are required to maintain precision during close encounters. We conduct N-body simulations using the REBOUND package \citep{rein_rebound_2012} and integrate them with the IAS15 integrator \citep{rein_ias15_2015}, which utilizes a high-order, adaptive time-stepping scheme. The integrator therefore automatically shortens time steps during periapsis passages of high‑eccentricity trajectories to resolve close approaches with high fidelity, while maintaining a rigorous constraint on global integration errors throughout the simulation.

\subsection{Definition of Protection}\label{subsec:define}

\textcolor{black}{We run each set of simulations until one of the following stopping criteria is met: (1) the asteroid is either ejected from the inner system or collides with one of the massive bodies (the white dwarf or the planet); or (2) the simulation time reaches the upper limit of $10\,T_{\mathrm{ast}}$. In cases where multiple criteria might be satisfied simultaneously, the earliest interaction is recorded. }Each asteroid is then classified into one of \textcolor{black}{five} mutually exclusive outcome classes based on three diagnostic variables: (i) the minimum distance between the asteroid and the white dwarf, \(d_{\min}\); (ii) the asteroid's final orbital energy, \(E_{\mathrm{final}}\) (negative values indicate bound orbits); \textcolor{black}{and (iii) the minimum distance between the asteroid and the planet, \(d_{p,\min}\).} The outcome classes are defined as:
\begin{spacing}{1.05}
\begin{description}
\item[\textbf{Ejected‑OutRoche (Ejected‑OR)}] \((E_{\mathrm{final}}>0,\; d_{\min}>r_{\mathrm{Roche}},\; d_{p,\min}>R_{\mathrm{p}})\): ejected before entering the Roche radius and without colliding with the planet.
\item[\textbf{Ejected‑InRoche (Ejected‑IR)}] \((E_{\mathrm{final}}>0,\; R_{\mathrm{WD}}<d_{\min}<r_{\mathrm{Roche}},\; d_{p,\min}>R_{\mathrm{p}})\): ejected after passing inside the Roche radius but prior to any impact.
\textcolor{black}{\item[\textbf{Collide}] \((d_{p,\min}<R_{\mathrm{p}})\): collides with the close‑in planet.}
\item[\textbf{Crash}] \((d_{\min}<R_{\mathrm{WD}}, d_{p,\min}>R_{\mathrm{p}})\): crash into the white dwarf.
\item[\textbf{Bound}] \((E_{\mathrm{final}}<0,\; d_{\min}>R_{\mathrm{WD}},\; d_{p,\min}>R_{\mathrm{p}})\): remains gravitationally bound to the white dwarf at the end of the integration and does not collide with the planet or the white dwarf.
\end{description}
\end{spacing}

\begin{figure*}[ht!]
\includegraphics[width=\textwidth]{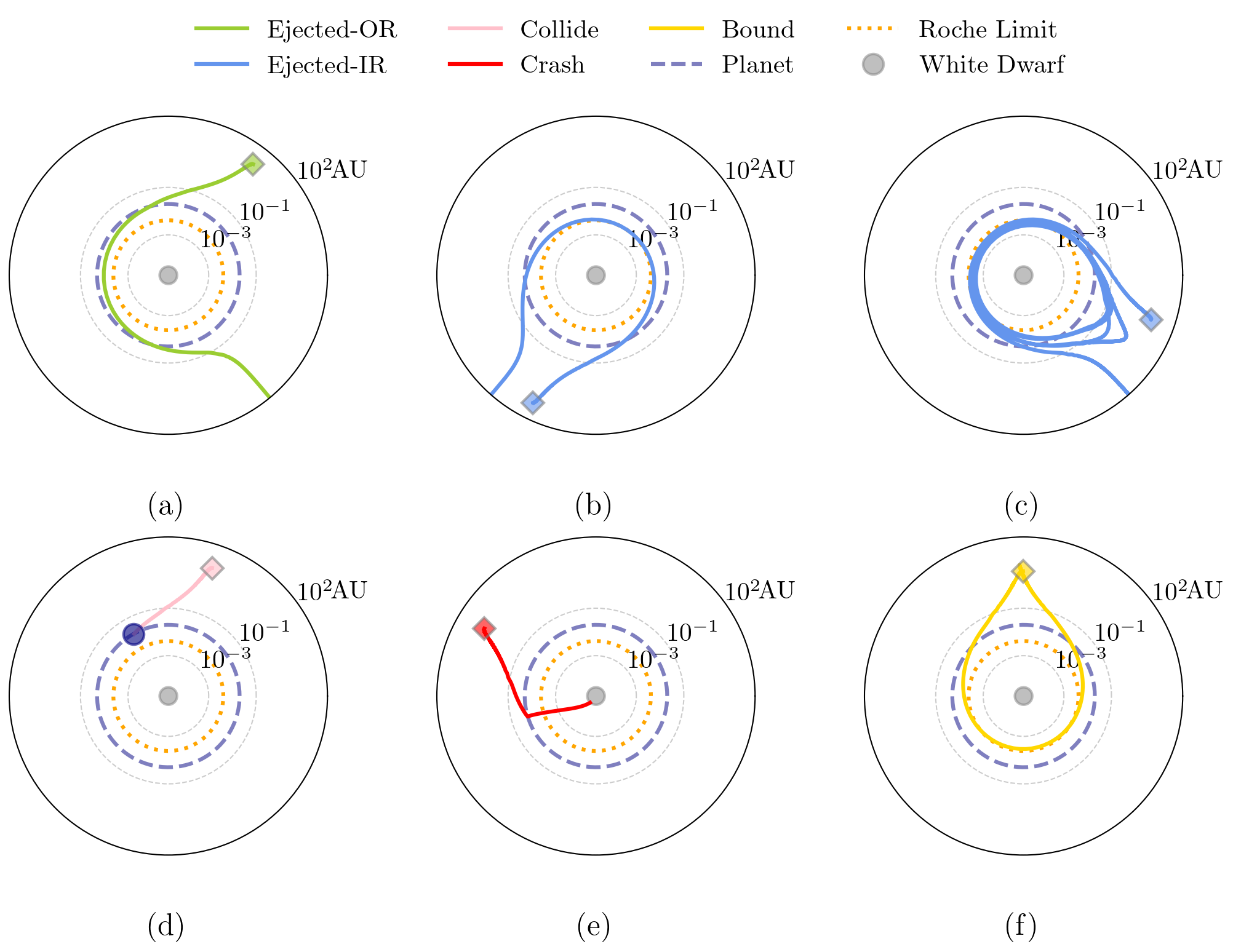}
\caption{Typical asteroid trajectories of four classifications in section \ref{subsec:define} on a logarithmic scale, with units in au. Line colors denote classifications and the diamond markers indicate the starting points of the asteroids. (a) belong to ``Ejected-OutRoche": The asteroid was ejected before entering into the roche limit; (b) and (c) are ``Ejected-InRoche": They are both ejected after entering into the roche limit, the difference is the duration they spent within the Roche limit. (d) shows the ``Collide" case: The asteroid collided with the planet when crossing the planetary orbit (We depict the planet at the moment of collision (navy circles); note that the planetary radii do not represent the planets' actual sizes.). (e) shows the ``Crash" case: The asteroid was scattered directly to the WD. And (f) are still ``Bound": They are still orbiting the white dwarf until the end of simulation.}\label{fig:orbits}
\end{figure*}

Without close-in orbit planets, all bodies whose initial orbits cross the Roche limit would pollute the white dwarf. The existence of planets will change asteroids' fate, and we quantify the protective effect of planets with the proportion of asteroids that are finally ejected at the end of the simulation and that collide with the planet. We define the protective fraction
\begin{equation}
F_{\mathrm{Protection}}=F_{\mathrm{Ejected-OR}}+F_{\mathrm{Ejected-IR}}\textcolor{black}{+F_{\mathrm{Collide}}}\label{equa:protection}
\end{equation},
where $F_{\mathrm{Ejected-OR}}$ and $F_{\mathrm{Ejected-IR}}$ are the fractions of asteroids in the \textit{Ejected‑OutRoche} and \textit{Ejected‑InRoche} classes, respectively, and $F_{\mathrm{Collide}}$ are the fraction in the \textit{Collide} class. 

 \begin{figure*}[ht]
     \centering
     \includegraphics[width=1\textwidth]{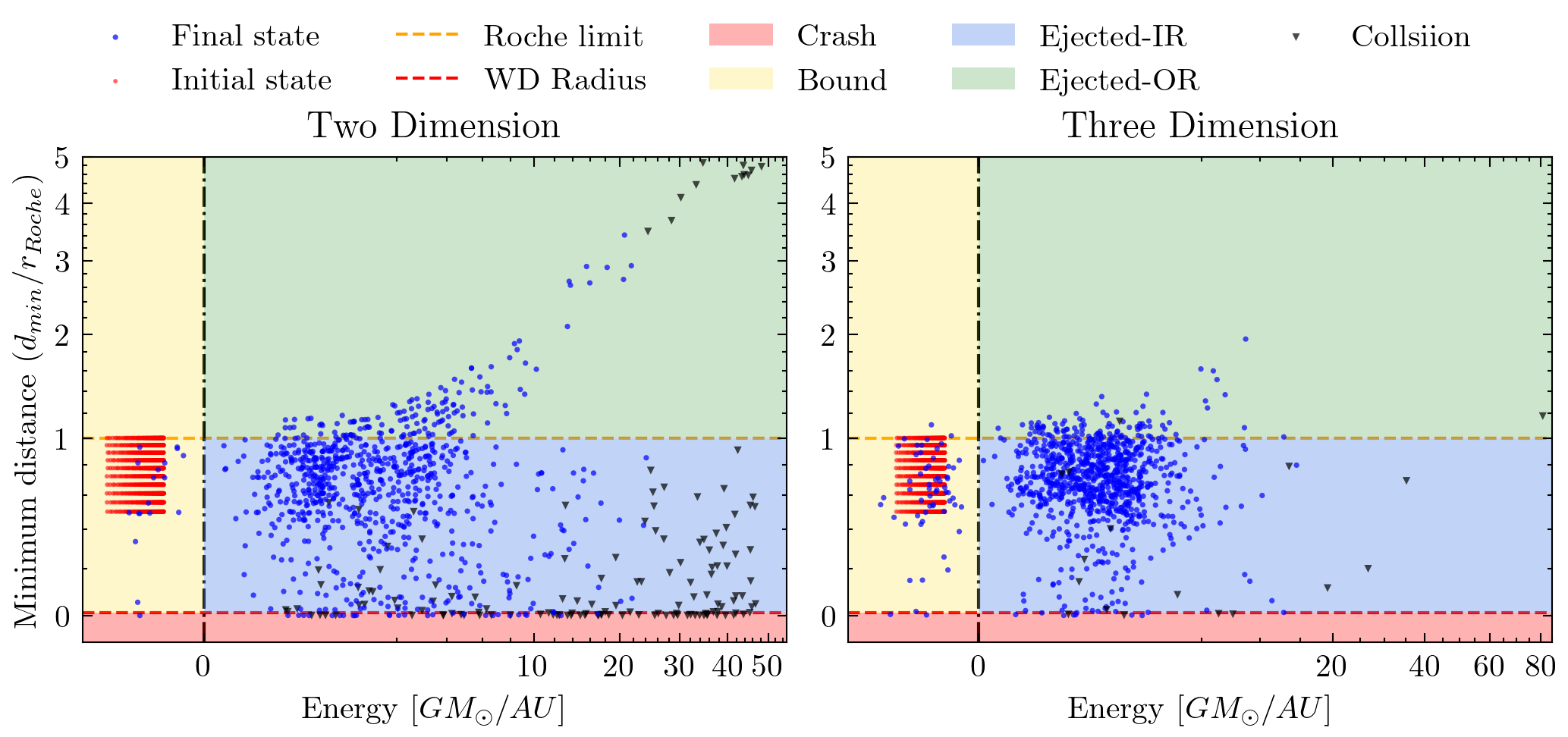}
     \caption{Distribution of asteroids' final energy and minimum distance to the WD during the entire simulation. The system parameters selected for this simulation are the benchmark parameters of WD1856+534b system given in Table.~\ref{tabel:parameter}. The red dots in the figure represents the initial state of asteroids. black dots represent the final state of each asteroid group at the end of the simulation; the color of the region where they land denotes their ultimate fate. And the black inverted triangles indicate those that collided with planets. 
     }\label{fig:distribution}
 \end{figure*}
 
\section{Results}\label{sec:results}
\subsection{Typical cases}\label{subsec:typical}

\begin{table*}[ht]
\centering
\caption{The fractions of asteroids within four outcome categories and the effects of protection under different planetary parameter combinations}\label{tab:population}
\renewcommand{\arraystretch}{1.25}
\begin{tabular}{l c c c c c c c r}
\toprule
\toprule
\multirow{2}{*}{\textbf{Condition}} &
\multicolumn{2}{c}{\textbf{Planet parameter}} &
\multicolumn{5}{c}{\textbf{Population}} &
{\textbf{Protection}} \\
\cmidrule(lr){2-3} \cmidrule(lr){4-8} \cmidrule(lr){9-9}
 & $m_p\,[M_\mathrm{Jup}]$ & $a_p\,[\mathrm{au}]$ &
 $F_{\mathrm{Ejected-OR}}$ &
 $F_{\mathrm{Ejected-IR}}$ &
 $F_{\mathrm{Bound}}$ &
 $F_{\mathrm{Crash}}$ &
 $F_{\mathrm{Collide}}$ &
 $F_{\mathrm{Protection}}$ \\
\midrule
\midrule
\multicolumn{9}{l}{\textbf{Two dimension}}\\
\midrule
Benchmark case & 13.8 & 0.02 & 12.0\% & 69.4\% & 1.0\% & 6.8\% & 10.8\% & 92.2\% \\
\midrule
\multirow{3}{*}{Change $m_p$} 
& 5    & 0.02 & 6.8\%  & 78.1\% & 1.5\% & 2.9\% & 10.7\% & 95.6\% \\
& 1    & 0.02 & 1.6\%  & 71.3\% & 9.9\% & 0.0\% & 17.2\% & 90.1\% \\
& 0.1  & 0.02 & 0.4\%  & 64.0\% & 17.8\% & 0.0\% & 17.8\% & 82.2\% \\
\midrule
\multirow{3}{*}{Change $a_p$}  
& 13.8 & 0.1  & 24.4\% & 53.0\% & 2.3\% & 16.6\% & 3.7\% & 81.1\% \\
& 13.8 & 0.5  & 27.6\% & 38.3\% & 11.9\% & 20.8\%  & 1.4\%  & 67.3\% \\
& 13.8 & 1.0  & 21.3\% & 30.2\% & 25.4\% & 20.9\%  & 2.2\%  & 53.7\% \\
\midrule                              
$[1.0,2.0]r_{\mathrm{Roche}}$ & 13.8 & 0.02 & 66.2\%  & 17.7\% & 0.6\% & 5.6\% & 9.9\% & 93.8\% \\
\midrule 
\midrule 
\multicolumn{9}{l}{\textbf{Three dimension}}\\
\midrule
Benchmark case & 13.8 & 0.02 & 7.65\%  & 88.4\% & 1.19\% & 1.36\% & 1.4\% & 97.39\% \\
\midrule
\multirow{3}{*}{Change $m_p$} & 5 & 0.02 & 4.8\%  & 88.9\% & 3.5\% & 0.8\% & 2.0\% & 95.7\% \\
 & 1 & 0.02 & 0.6\%  & 66.9\% & 30.2\% & 0.2\% & 2.1\% & 69.6\% \\
 & 0.1 & 0.02 & 0.1\%  & 4.4\% & 94.9\% & 0.0\% & 0.6\% & 5.1\% \\
 \midrule
\multirow{2}{*}{Change $a_p$} & 13.8 & 0.1 & 14.0\%  & 74.5\% & 9.5\% & 1.8\% & 0.2\% & 88.7\% \\
& 13.8 & 0.5 & 17.2\%  & 39.9\% & 40.7\% & 2.2\% & 0.0\% & 57.1\% \\
& 13.8 & 1.0 & 11.6\%  & 20.1\% & 66\% & 2.3\% & 0.0\% & 31.7\% \\
\midrule 
\midrule 
{\textbf{Fragmentation}}\\
\midrule
$r_\mathrm{frag}=10~\mathrm{km}$ & 13.8 & 0.02 & 12.0\%  & 69.72\% & 1.16\% & 7.24\% & 9.88\% & 91.6\% \\
$r_\mathrm{frag}=50~\mathrm{km}$ & 13.8 & 0.02 & 12.0\%  & 69.76\% & 1.22\% & 7.28\% & 9.74\% & 91.5\% \\
$r_\mathrm{frag}=100~\mathrm{km}$ & 13.8 & 0.02 & 12.0\%  & 69.04\% & 1.22\% & 7.14\% & 10.6\% & 91.64\% \\
$r_\mathrm{frag}=500~\mathrm{km}$ & 13.8 & 0.02 & 12.0\%  & 69.42\% & 1.10\% & 7.48\% & 10.0\% & 91.42\% \\
\bottomrule
\bottomrule
\end{tabular}
\renewcommand{\arraystretch}{1.0}
\end{table*}

In the previous section, we established criteria to assess planetary influence on small bodies and classify outcomes into five distinct classes. Below, we analyze each class and present representative examples.

Fig.\ref{fig:orbits}(a) illustrates the first class (\textit{Ejected‑OutRoche}), in which planetary perturbations deflect asteroid trajectories before they reach the Roche sphere. This deflection confines these asteroids exterior to the planet’s orbit, preventing close approaches to the white dwarf and representing the strongest form of planetary protection.

In the second outcome class (\textit{Ejected‑InRoche}), asteroids can be expelled either by a single close planetary encounter or via multiple encounters that progressively increase their orbital energy, ultimately producing unbound trajectories. Representative examples are shown in Fig. \ref{fig:orbits} (b) (c). Although classified as the same category, asteroids with different initial conditions exist within the Roche limit of a white dwarf for significantly varying durations. Because tidal disruption is not instantaneous upon entering the outer Roche limit \citep{malamud_tidal_2020}, these residence‑time differences produce divergent evolutionary outcomes. Bodies with short Roche‑limit transits ($\Delta t  \lesssim \tau_\mathrm{disrupt}$) may be ejected before significant tidal fragmentation, thereby avoiding accretion onto the white dwarf. Conversely, prolonged residence ($\Delta t \gg \tau_\mathrm{disrupt}$) leads to full tidal disintegration, producing debris that can ultimately pollute the white dwarf atmosphere. This outcome represents an intermediate, imperfect protective effect—termed ``weak protection"—in which the planet reduces but does not eliminate the risk of white‑dwarf pollution. The population of this category can provide a wide reference constraint on the ``protection" effect of planets. Later in section~\ref{subsec:fragment} we address a critical extension: whether planets can effectively clear debris prior to accretion when asteroids are shattered inside the Roche limit.

\textcolor{black}{When an asteroid's trajectory is strongly perturbed by planetary gravity, it may undergo a close encounter that leads to a direct impact with the planet. Fig.~\ref{fig:orbits}(d) illustrates a typical example of such a collision. The orbit shows a sudden termination at the planetary radius. Once captured by the planetary atmosphere, these "pollutants" are effectively removed from the system, preventing any further interaction with the white dwarf. This direct interception constitutes an additional, efficient mechanism for planetary protection. }

Planetary scattering does not always redirect small bodies outward; in some cases it instead promotes inward scattering that can lead to white‑dwarf accretion. Fig.\ref{fig:orbits} (e) shows a representative trajectory in which a small body is scattered on a nearly radial path toward the white dwarf. We do not model post‑Roche processes (e.g., tidal fragmentation, collision cascades, and subsequent debris evolution) in the present simulations. However, including tidal‑disruption physics could alter this outcome: fragments produced inside the Roche limit may still be cleared by planetary perturbations before accretion, thereby reducing the pollution. We will present the results later in section ~\ref{subsec:fragment}.
 
\textcolor{black}{Fig.\ref{fig:orbits} (f) present a representative bound case ($E_\mathrm{final}<0$). Such orbits often arise from planetary perturbations that significantly increase orbital periods, thereby delaying the asteroid's return to the white dwarf's vicinity.Unless a planet ejects a body in a single scattering event, the object typically undergoes multiple close encounters, each of which can lead to either ejection or inward scattering. While the cumulative probability of ejection increases with each encounter \citep{oconnor_pollution_2023}, asteroids in systems with low-mass planets may persist for thousands of years, potentially crossing the Roche limit numerous times. Since an asteroid could be tidally disrupted and accreted before ejection occurs, we classify these temporarily bound objects as a separate category and exclude them from the protection fraction calculation to ensure a conservative estimate of the planet's defensive role.}

\subsection{Population analysis}\label{subsec:population}

While the previous section categorized individual asteroid trajectories, this section provides a statistical quantification of these outcomes across a broad parameter space. We evaluate the effectiveness of planets as dynamical shields by analyzing the relative proportions of each outcome class. The comprehensive statistical results for various planetary mass and orbital configurations are summarized in Table~\ref{tab:population}.

\subsubsection{Two dimension}

Although the \textit{Ejected-OutRoche} category represents the most effective form of planetary protection—where asteroids are removed before they can ever threaten the white dwarf—it is not the dominant outcome. In our benchmark case (see Table~\ref{tabel:parameter}), only 12\% of asteroids are deflected while remaining outside the Roche limit; these are visualized as the blue points within the black-shaded region in the left panel of Fig.\ref{fig:distribution}. \textcolor{black}{The statistics in Table~\ref{tab:population} reveal that at a fixed semi-major axis of 0.02au, the fraction of asteroids achieving this fate decreases significantly as planetary mass declines, reflecting a reduced scattering cross-section. Conversely, for a fixed mass of $13.8~M_\mathrm{Jup}$, the proportion of this category increases slightly as the planetary semi-major axis is extended beyond 0.1~au, exceeding 20\%. This suggests that more distant planets, while having a smaller overall impact area, may more gently "nudge" asteroids into unbound orbits before they reach the inner system.}

In contrast, the \textit{Ejected-InRoche} class constitutes the most populous outcome, where asteroids are scattered out of the system only after penetrating the Roche limit. In our benchmark case, this class accounts for approximately 70\% of the total population. Such a high fraction implies that while planets are efficient at eventually clearing eccentric asteroids, they often fail to prevent the initial penetration into the white dwarf's immediate vicinity. \textcolor{black}{This proportion shows a non-monotonic dependence on planetary mass: it peaks at nearly 80\% as the mass decreases slightly from the benchmark, but subsequently declines for planets significantly smaller than one Jupiter mass, where the scattering efficiency drops. Furthermore, the fraction of this class decreases as the planetary semi-major axis increases, as more distant planets exert a weaker dynamical influence on the inner regions near the Roche limit.} Since this category includes asteroids that may have completed multiple periastron passages, we will further quantify the planet's impact on the resulting tidal fragments in section~\ref{subsec:fragment}.

\begin{figure}[ht]
\includegraphics[width=0.48\textwidth]{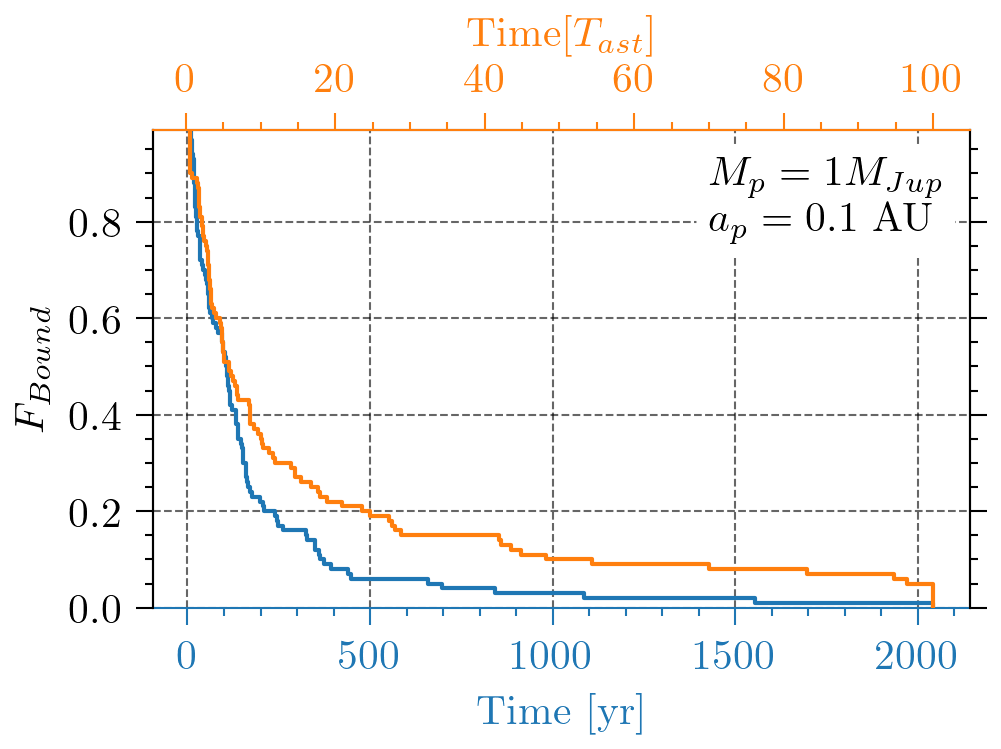}
\caption{The proportion of bound asteroids to all simulated asteroids as a function of simulation time. The two different colors represent the trends under different time units.\label{fig:bound}}
\end{figure}

Asteroids remaining \textit{Bound} to the system at the end of our simulations are the primary potential sources for white dwarf pollution. For planets located at 0.02~au, we find that when the mass exceeds $5~M_\mathrm{Jup}$, the bound fraction remains below 2\%, indicating high clearing efficiency. \textcolor{black}{However, $F_\mathrm{Bound}$ increases significantly for lighter or more distant planets. To illustrate the long-term clearing process, Fig.~\ref{fig:bound} shows the temporal evolution of the bound fraction for a $1~M_\mathrm{Jup}$ planet at $0.1~\mathrm{au}$. The planet requires approximately 100 initial orbital periods—roughly 2,000~years—to fully clear this population. This delay is critical, as it provides a window for asteroids to undergo tidal disruption or radiative circularization \citep{li_accretion_2021}, potentially leading to accretion before dynamical ejection can take place.}

A small but distinct subset of asteroids undergoes more violent dynamical evolution, leading to the \textit{Crash} outcome. In our benchmark case, approximately 7\% of the planet-influenced asteroids are scattered directly onto the white dwarf’s surface. \textcolor{black}{This fraction is highly sensitive to planetary parameters: at a fixed semi-major axis, it drops to 0\% for a $1~M_\mathrm{Jup}$ planet, but climbs to 20\% as the planet's distance increases beyond 0.5 au. While such direct impacts represent the most immediate form of pollution, they differ from the gradual accretion typically inferred from debris disks. A high \textit{Crash} fraction suggests a regime where planetary perturbations are strong enough to bypass tidal disruption entirely. If a reservoir of eccentric objects is continuously replenished, this direct channel—even at a 10\%--20\% efficiency—could contribute to the observed steady-state pollution, albeit with different temporal characteristics than the disk-fed accretion model.}

\textcolor{black}{Beyond gravitational scattering, the physical size of giant planets allows them to act as a direct ``sink" for infalling debris. In our benchmark case, approximately 10\% of the asteroids end their evolution by colliding with the planet. Interestingly, due to the mass-radius relationship for giant planets Eq.~(\ref{equa:radius}), planets with masses around 0.1 to 1 $M_\mathrm{Jup}$ possess larger physical cross-sections than their more massive counterparts. Consequently, these intermediate-mass planets exhibit higher collision fractions, making them more effective at ``sequestering" potential polluters. This suggests that the chemical signatures of pollution in some white dwarf systems might be mirrored by the enrichment of their close-in giant planets, offering a potential ``planetary shield" signature for future atmospheric characterization. However, this protective mechanism becomes inefficient as the planet's semi-major axis increases beyond 0.5 au, where the collision fraction falls below 5\%. This decline highlights that for distant planets, protection is achieved almost entirely through dynamical ejection rather than physical interception.}

\textcolor{black}{For asteroids that do not initially cross the Roche limit, the presence of a close-in planet provides an additional layer of security. Our results show that the majority of such objects are ejected from the system before they can even reach the Roche zone. Specifically, in the benchmark case, only about a quarter of the asteroids are driven into the Roche limit by planetary perturbations. Among these, 17.7\% are subsequently ejected after their inward excursion (\textit{Ejected-InRoche}), and only a minimal fraction (5\%) result in a direct \textit{Crash} onto the white dwarf. Notably, the collision fraction remains largely insensitive to the asteroids' initial perihelion locations. This suggests that the planetary ``shield" is not merely a local obstacle but extends its influence well beyond the immediate tidal disruption zone, efficiently ``cleaning" the system of potential polluters before any significant inward orbital migration—such as that driven by radiative effects—can occur.}

\subsubsection{Three dimension}

Although planets and small bodies in the Solar System are predominantly coplanar, many exoplanetary systems host planets with large eccentricities and significant mutual inclinations, enabling pollutant delivery from a wide range of directions. In addition, contaminants originating in the outer reservoir (e.g., exo‑Oort‑cloud comets) may approach the inner system from arbitrary inclinations. To compare the difference of planetary influence under coplanar and non-coplanar circumstances, we randomly selected 1,000 simulations and plotted the distribution of asteroids in the energy–distance plane (Fig.~\ref{fig:distribution}) 

The probability of extremely close planet–asteroid encounters is higher for coplanar geometries than for three‑dimensional configurations, producing stronger instantaneous perturbations in the coplanar case. As shown in Fig.~\ref{fig:distribution}, the distribution of outcomes in the energy–distance plane is broader for the coplanar sample than for the three‑dimensional sample. Additionally, we quantify the difference in coplanar and non-coplanar scenarios by the average minimum gravitational interaction distance between the asteroid and the planet. In the non‑coplanar case this distance is $2.8~R_\mathrm{Hill}$, while in the coplanar case it is $2.1~R_\mathrm{Hill}$. At this interaction distance, the influence of the planet to alter the fates of the asteroids in three dimension remains significant, though the magnitude of individual changes in orbital direction decreases. Accordingly, the fraction of \textit{Ejected‑OutRoche} objects ($F_{\mathrm{Ejected-OR}}$) falls to about 8\% in the three‑dimensional experiments. Simultaneously, the \textit{Ejected‑InRoche fraction} ($F_{\mathrm{Ejected-IR}}$) increases to roughly 88\%, so the ejection fraction in three dimensions exceeds that found for coplanar runs. The fraction of direct collisions with either white dwarfs or planets has fallen to about 1\%. Consequently, the total protective fraction, $F_{\mathrm{Protection}}$, rises to approximately 97\% for isotropic pollutant sources. 

\textcolor{black}{However, this superior protective effect in three dimensions is highly sensitive to planetary parameters. While the shield remains stronger in 3D for slight increases in semi-major axis or small decreases in mass, it is not universally more effective than the coplanar shield. As the planetary mass continues to decrease or the semi-major axis increases, the effectiveness of the 3D protection degrades much more precipitously than its 2D counterpart. Consequently, the protection efficiency in the three-dimensional scenario suffers a far more dramatic reduction than in the two-dimensional case. For instance, when the planetary mass is reduced to $0.1~M_\mathrm{Jup}$ at 0.02~au, the 3D protective effect nearly vanishes, dropping to only 5.1\%, while the 2D protection remains robust at 82.2\%. }

\subsection{Effect of fragmentation\label{subsec:fragment}}
In the preceding simulations, asteroids were treated as intact particles even after crossing the Roche limit. However, a more realistic and observationally motivated scenario is that such bodies undergo tidal disruption upon close approach, forming a debris stream or disk that eventually accretes onto the white dwarf \citep{veras_evolution_2024}. Since the single-body approximation cannot capture the complex dynamical evolution that occurs within the Roche limit, we extend our model to incorporate tidal fragmentation. This allows us to evaluate whether the divergent trajectories of fragments, under the combined influence of the white dwarf and the planetary perturbations, significantly alter the overall protection efficiency.

\begin{figure}[htb!]
\includegraphics[width=0.48\textwidth]{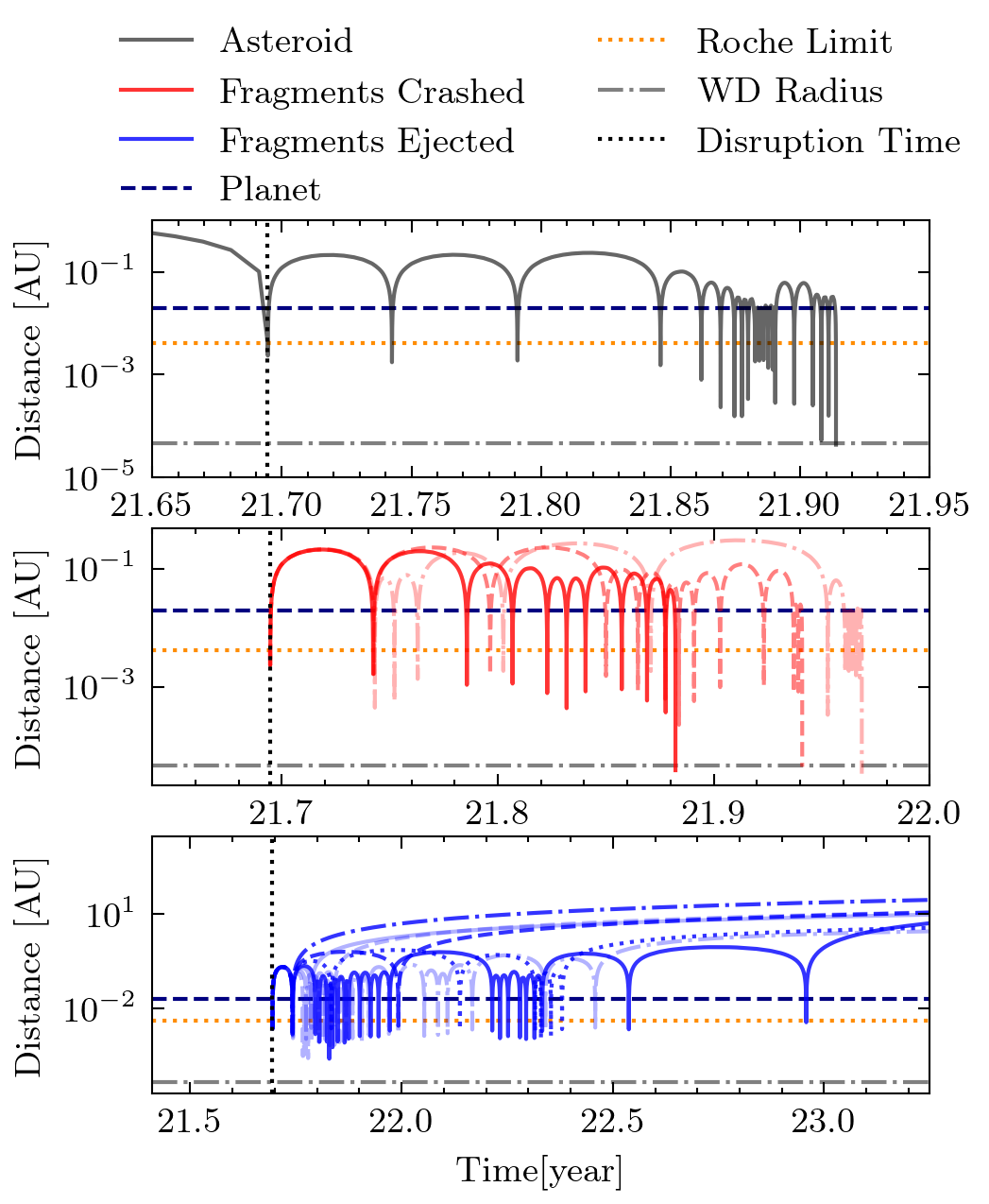}
\caption{The evolution of the distance between the particles and the white dwarf over time. The top panel show the evolution of asteroid if tidal disruption is ignored. The middle and bottom panel show the evolution of the fragments considering the tidal disruption, with different line styles representing different fragments.}\label{fig:fragments_case}
\end{figure}

\begin{figure*}[htb!]
\includegraphics[width=1\textwidth]{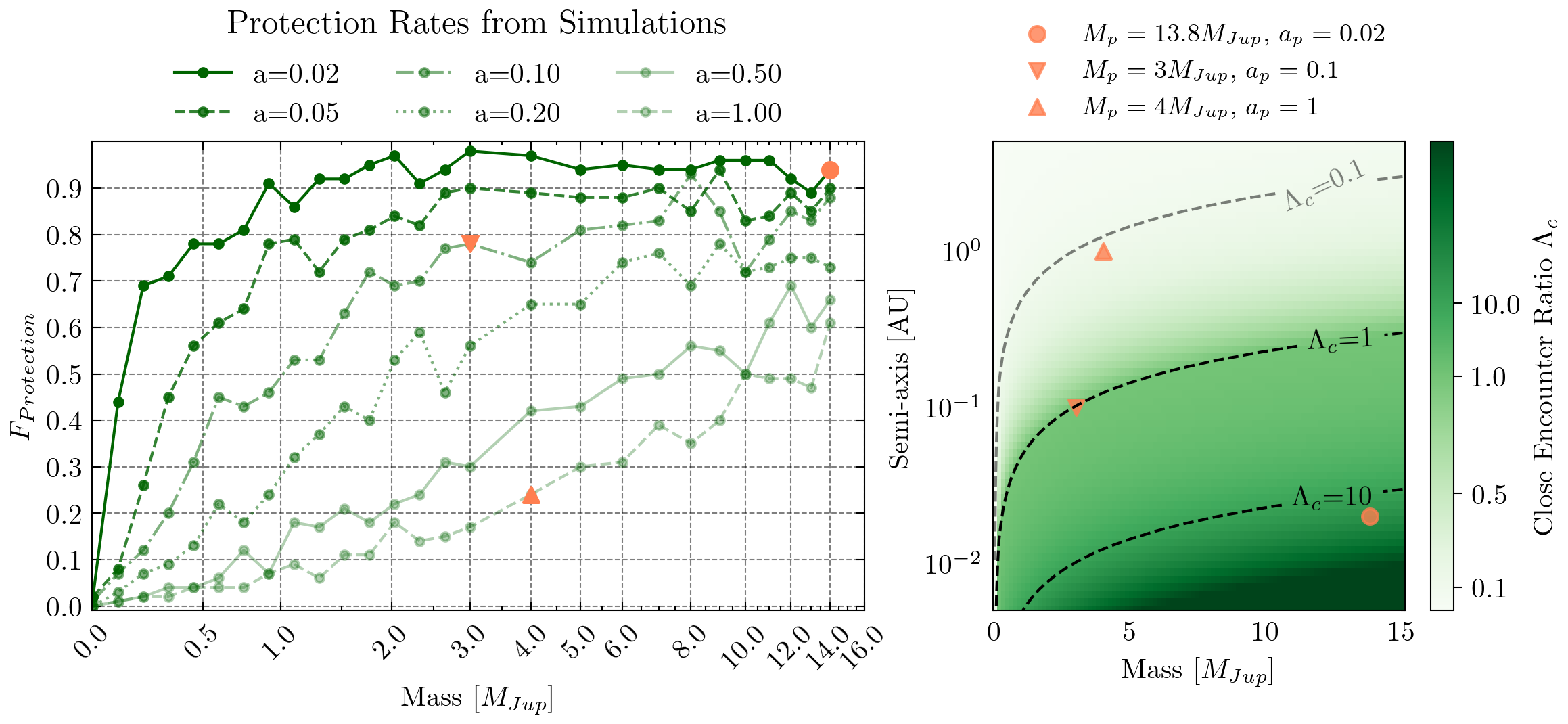}
\caption{Left panel: Protection rates ($F_{protection}$) given by Eq.~\eqref{equa:protection} as a function of mass, with different line styles representing distinct semi-major axis ($a_p$) values. Right panel: Mapping the modified close encounter ratio (Eq.~\eqref{equa:safro}) in the planetary mass-semi major axis diagram.}\label{fig:planet}
\end{figure*}

Following the analytical framework of \citet{malamud_tidal_2020}, we adopt a simplified disruption model based on the following assumptions: (1) The parent asteroid is assumed to be spherical and dynamically undisturbed prior to entering the Roche limit. (2) Upon the first crossing of the Roche limit, the body disrupt instantaneously into 10 fragments, each subsequently treated as a massless test particle. (3) These fragments inherit the instantaneous center‑of‑mass velocity of the parent body at the moment of disruption, while their initial positions are distributed isotropically at a distance $r_\mathrm{frag}$ from the parent's center. In this prescription, the divergent evolution of fragments is driven solely by their spatial offsets, which translate into slight differences in their orbital energy and angular momentum relative to the white dwarf. 

\textcolor{black}{The spatial offset $r_\mathrm{frag}$, which scales with the asteroid's physical radius, introduces a degree of dynamical diversity among fragments. Specifically, there exists a critical radius $r_\mathrm{crit}$ above which fragments with $r_\mathrm{frag}>r_\mathrm{crit}$ become gravitationally unbound from the white dwarf immediately upon disruption. Using Equation (2) from \citet{malamud_tidal_2020}, we calculate the critical value for our system:
\begin{equation}
\centering
r_\mathrm{crit}=\frac{r_\mathrm{Roche}^{2}}{2a_\mathrm{ast}-r_\mathrm{Roche}^{2}}\approx127~\mathrm{km}.
\end{equation}
In the absence of a planet, fragments that become unbound due to tidal forces alone would be unable to pollute the white dwarf. However, since our definition of planetary protection ($F_\mathrm{Protection}$) is predicated on a 100\% baseline pollution rate, we primarily stipulate $r_\mathrm{frag}<r_\mathrm{crit}$ to ensure that any observed protection is attributable to planetary influence. Nevertheless, considering that realistic polluters may be larger \citep{veras_evolution_2024}, we also explore cases where $r_\mathrm{frag}>r_\mathrm{crit}$ as a reference. Consequently, we tested $r_\mathrm{frag}$ values of 10 km, 50 km, 100 km, and 500 km, with the corresponding statistical outcomes presented in Table~\ref{tab:population}.}

\textcolor{black}{We conducted these fragmentation simulations using exactly the same parameters as the 2D benchmark case to isolate the effect of tidal disruption on the resulting statistics.} Notably, the fraction of asteroids ejected prior to reaching the Roche limit (\textit{Ejected-OR}) remains identical to the single-body model, as only those parent bodies that penetrate $r_\mathrm{Roche}$ are subject to fragmentation. Upon disruption, however, fragments from the same parent can follow markedly divergent dynamical trajectories. To illustrate this stochasticity, Fig.~\ref{fig:fragments_case} presents a representative case comparing the evolution of a parent asteroid and its child fragments. In the single-body integration (upper panel), the parent body is directly accreted by the white dwarf. In contrast, the lower panels reveal that while 3 out of 10 fragments (30\%) follow the parent’s fate and accrete onto the white dwarf, the remaining 7 fragments (70\%) are eventually ejected from the system due to subsequent planetary perturbations. This example highlights how tidal disruption can transform a potential accretion event into a series of ejections, diversifying the possible outcomes for the original pollutant mass.


 As discussed in section~\ref{subsec:typical}, asteroids in the \textit{Ejected-IR} category exhibit a wide diversity in their encounter histories prior to ejection. We now refine our analysis of this variation by tracing the ultimate fates of individual fragments. By assigning each fragment a mass fraction of 0.1 and adopting a fragment-weighted accounting scheme, we can determine the ejection and accretion fractions with higher precision than by treating parent bodies as indivisible units. For instance, as illustrated in Fig.~\ref{fig:fragments_case}, a parent body previously categorized as a \textit{Crash} (count = 1) is now recorded as contributing 0.7 to the \textit{Ejected-IR} fraction and 0.3 to the \textit{Crash} fraction. This refined approach effectively captures how the stochastic history of close planetary encounters dictates the divergent evolutionary paths of an asteroid’s constituent debris.
 
 Under this fragment-weighted accounting scheme, the overall protective fraction is $F_{\mathrm{Protection}}\approx 91.5\%$. This value is remarkably consistent with the $F_{\mathrm{Protection}}=92.2\%$ reported for the single‑asteroid model (Table.~\ref{tab:population}, first row), indicating that fragmentation does not fundamentally undermine the planetary shield. \textcolor{black}{Crucially, the initial radius of the asteroid has a negligible effect on the final outcome. The insensitivity of the protection fraction to the asteroid radius (from 10 km to 500 km) suggests that the phase space for ejection is dominated by the planet's strong scattering. The initial 'energy spread' caused by tidal disruption—even for a 500 km body—is insufficient to nudge fragments out of the planet's vast gravitational reach or to significantly alter the statistical probability of accretion versus ejection.} Although this simplified fragment model does not capture all real‑world complexities, it provides a useful first‑order reference: Eq.~\eqref{equa:protection} yields a reasonable estimate of the planetary protection efficiency.
 
\subsection{Effect of planet mass and orbital parameter\label{subsec:planet}}
In this subsection, we investigate how planetary mass and semi-major axis affect protective efficiency, identifying the conditions under which planets provide significant pollution shielding

Our simulations sample planetary masses $M_p$ from 1 $M_{\mathrm{Earth}}$ to 14 $M_{\mathrm{Jup}}$ and semi-major axes $a_p$ from 0.02 au to 1 au. For each $(M_p,a_p)$ pair, we perform 100 simulations to obtain outcome distributions analogous to those in Section~\ref{subsec:population}. Table~\ref{tab:population} summarizes representative results for selected $(M_p,a_p)$ combinations and Fig.~\ref{fig:planet} illustrates the dependence of $F_\mathrm{Protection}$ on planetary mass across five fixed semi-major axes.

The key findings are as follows. First, protective efficiency scales positively with planetary mass, approaching a maximum of over 90\%. For planets at 0.02 au, masses $\gtrsim 1M_\mathrm{Jup}$ prevent more than 90\% of potential asteroid incursions, demonstrating the formidable shielding power of close-in gas giants. Second, \textcolor{black}{the overall shielding effect of the planet diminishes with increasing semi-major axis.} At $a_p=0.5~\mathrm{au}$ the protective fraction drops by $\gtrsim 20\%$ across the explored mass range. Beyond $a_p=1~\mathrm{au}$, even the most massive planets in our sample eject only $\approx$ 60\% of asteroids. \textcolor{black}{This declining trend with distance stems from a competition between scattering strength and encounter frequency. Although the ratio of a planet's surface escape velocity to the local Keplerian velocity increases at larger distances—which theoretically favors ejection during a single close encounter—this is outweighed by the longer orbital periods and the significantly reduced frequency of encounters. Consequently, within a finite integration timescale, the cumulative probability of a successful ejection is lower for more distant planets, even if they are intrinsically ``stronger" scatterers per event.}

The direct scattering effect of a planet can be estimated analytically based on its mass and orbital parameters. If the kinetic energy imparted during a planetary encounter exceeds the particle's gravitational binding energy, the particle becomes unbound and is prevented from accreting onto the white dwarf \citep{oconnor_pollution_2023}. To evaluate this protective capability, we adopt a qualitative analytical approach. While \citet{oconnor_pollution_2023} primarily addressed perturbations from distant encounters, close-in planets are more likely to undergo strong, deep encounters with asteroids. Our simulations indicate that the typical closest-approach separation scales as approximately twice the Hill radius ($2~R_{\mathrm{Hill}}$). By adopting this characteristic encounter distance, we modify the scattering parameter $\Lambda$ (Eq.~39 of \citealt{oconnor_pollution_2023}) to define a close-encounter ratio $\Lambda_c$:
\begin{equation}
\begin{aligned}
    \Lambda_{c} &=(\frac{GM_{p}}{2R_{\mathrm{Hill}}})/(\frac{GM_{\star}}{a_\mathrm{ast}})
    =(\frac{M_p}{M_{\star}})\frac{a_\mathrm{ast}}{2a_p(\frac{M_p}{3M_{\star}})^{\frac{1}{3}}}\\
    &\sim [\frac{M_p}{M_{Jup}}]^{\frac{2}{3}}[\frac{M_\star}{M_{\odot}}]^{\frac{2}{3}}[\frac{a_\mathrm{ast}}{10~\mathrm{au}}][\frac{a_p}{0.1~\mathrm{au}}]^{-1}.
\end{aligned}\label{equa:safro}
\end{equation}
The modified ratio of close encounter $\Lambda_c$ of a planet depends on the semi‑major axis $a_\mathrm{ast}$ of the interacting asteroid. Since our sampled asteroids cover a range of initial orbits, we adopt a representative midpoint of $a_\mathrm{ast}=5$ au for this estimation. The resulting values of $\Lambda_c$ for our simulated $(M_p,a_p)$ pairs are displayed in the right panel of Fig.~\ref{fig:planet}. Theoretically, planets with $\Lambda_c \gtrsim 1$ are able to produce strong direct‑scattering effects on asteroids. We highlight three representative cases in both panels of Fig.~\ref{fig:planet} to illustrate this correspondence. A filled circle marks the benchmark case discussed previously, where $\Lambda_c>10$. Two triangles indicate cases with $\Lambda_c \approx 0.1$ and $\Lambda_c \approx 1$, , which correspond to protection fractions of $\approx$80\% and $\approx$25\% in our simulations, respectively. Overall, the simulation outcomes show strong alignment with the theoretical expectations derived from the Safronov-like criterion, validating $\Lambda_c$ as a useful predictor of planetary shielding efficiency.

\section{Discussion\label{sec:discuss}}

\subsection{Influence of other parameters\label{subsec:influence}}

We first examined the influence of white dwarf mass by performing simulations using various mass estimates for WD~1856+534 reported in the literature. In these tests, planetary parameters were kept constant, and the integration timescale remained unchanged to ensure a controlled comparison. As shown in Table~\ref{tab:wdmass}, the resulting asteroid outcome fractions across different WD masses vary by no more than 0.5\%. This consistency confirms that the uncertainty in the white dwarf’s mass—within the observationally constrained range—does not significantly alter the dynamical evolution or the ultimate protective efficiency of the system.

\begin{table*}[htb!]
\centering
\caption{Distribution of asteroid outcomes and planetary protection fractions for different white dwarf mass estimates. *\textit{Note: While \citet{limbach_thermal_2025} constrained the planet's mass to 5.2 $M_{\mathrm{Jup}}$}, we use a fixed set of planetary parameters across all entries in this table to isolate the effect of stellar mass. These values are intended for sensitivity analysis rather than a direct fit to specific observations.}\label{tab:wdmass}
\renewcommand{\arraystretch}{1.25}
\begin{tabular}{l c c c c c c c r}
\toprule
\toprule
\textbf{White dwarf mass} & \textbf{Source} &
\textbf{Planet mass} &
\multicolumn{5}{c}{\textbf{Population}} &
{\textbf{Protection}} \\
 $M_\textrm{WD}\,[M_\odot]$& & $m_p\,[M_\mathrm{Jup}]$ &
 $F_{\mathrm{Ejected-OR}}$ &
 $F_{\mathrm{Ejected-IR}}$ &
 $F_{\mathrm{Bound}}$ &
 $F_{\mathrm{Crash}}$ &
 $F_{\mathrm{Collide}}$ &
 $F_{\mathrm{Protection}}$ \\
\midrule
0.518 & \citet{vanderburg_giant_2020} & 13.8 &  12.0\% & 69.4\% & 1.0\% & 6.8\% & 10.8\% & 92.2\% \\
0.576 & \citet{xu_geminigmos_2021} & 13.8 &  12.1\% & 70.5\% & 0.9\% & 6.5\% &  10.0\% & 92.6\% \\
0.605 & \citet{limbach_thermal_2025}*& 13.8 &  12.0\% & 71.1\% & 0.5\% & 6.8\% & 9.6\% & 92.7\% \\
\midrule
0.518 & \citet{vanderburg_giant_2020} & 5 &  5.5\% & 77.9\% & 2.2\% & 2.1\% & 12.3\% & 95.7\%\\
0.576 & \citet{xu_geminigmos_2021} & 5 &  4.7\% & 78.1\% & 2.7\% & 1.9\% & 12.6\% & 95.4\%\\
0.605 & \citet{limbach_thermal_2025} & 5 &  4.2\% & 77.2\% & 2.6\% & 1.9\% & 14.1\% & 95.5\% \\
\bottomrule
\bottomrule
\end{tabular}
\renewcommand{\arraystretch}{1.0}
\end{table*}

\begin{figure}[htb!]
     \centering
     \includegraphics[width=0.5\textwidth]{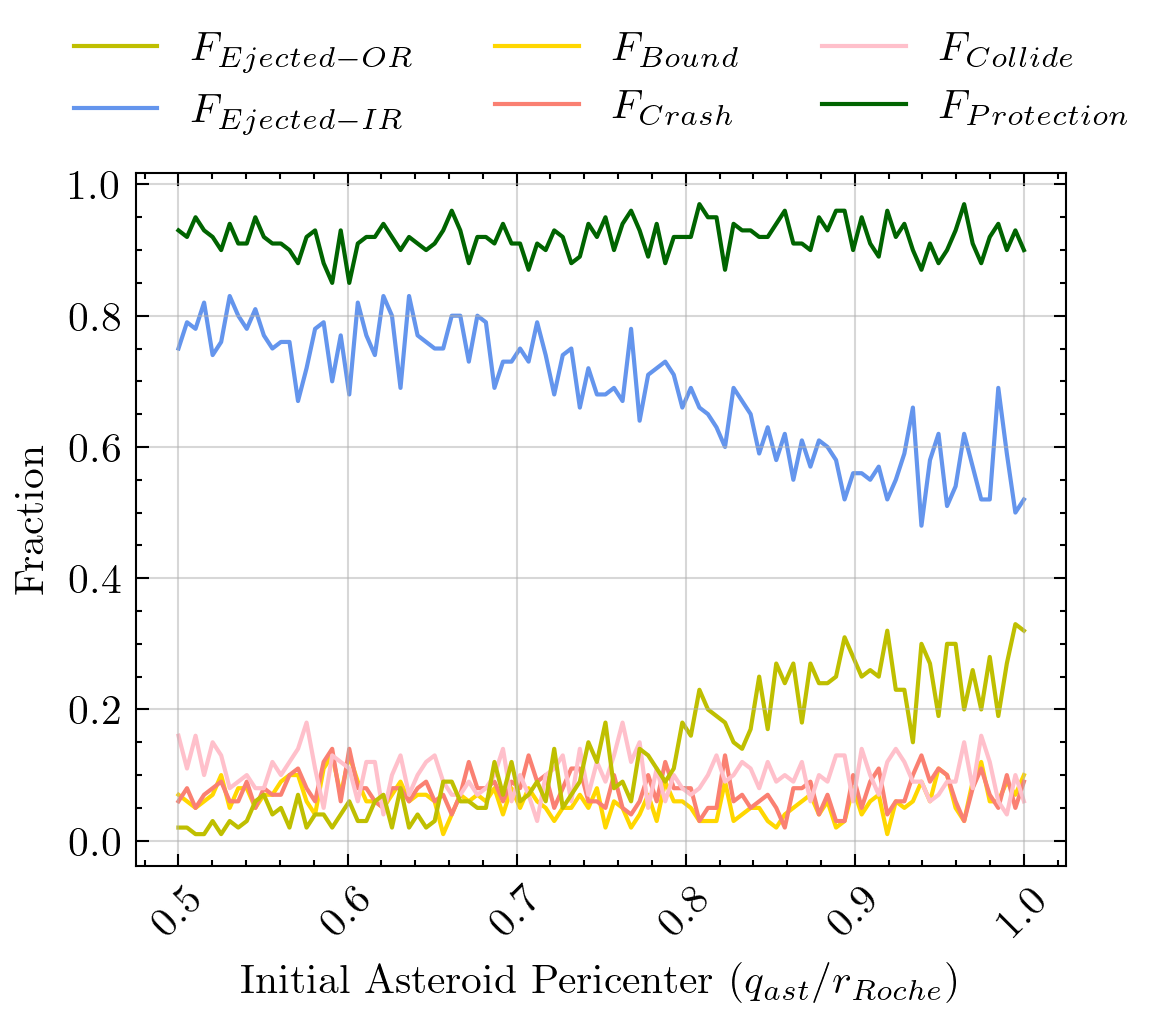}
     \caption{The relationship between the fractions of asteroids within classifications and the their initial pericenter.}\label{fig:astPara}
 \end{figure}

While the results presented thus far are based on a specific set of initial orbits, it is crucial to verify whether the inferred protective efficiency is sensitive to these assumptions. We therefore examine how varying the asteroids' initial semi-major axis ($a_\mathrm{ast}$) and periapsis ($q_\mathrm{ast}$) distributions alters the statistical outcomes. By employing a "double-scan" strategy—systematically varying $a_\mathrm{ast}$ while holding $q_\mathrm{ast}$ constant, and vice versa—we isolate the dynamical impact of each parameter. As shown in Figure~\ref{fig:astPara}, as the initial $q_\mathrm{ast}$ decreases, $F_\mathrm{Ejected-OR}$ declines, while $F_\mathrm{Ejected-IR}$ increases. This shift indicates that asteroids with smaller initial periapses are more likely to bypass the planet’s "outer guard" and reach the Roche limit. However, the combined $F_\mathrm{Protection}$ remains remarkably stable across the sampled $q_\mathrm{ast}$ range, confirming that the choice of initial periapsis has a negligible impact on the overall protective capacity. 

The dependence on the initial semi-major axis $a_\mathrm{ast}$ can be understood through Eq.~\eqref{equa:safro}. Theoretically, there exists a threshold $a_{in}$ above which the scattering ratio $\Lambda_c > 1$, signaling highly efficient ejection. For our benchmark case ($13.8~M_\mathrm{Jup}$ at $0.02~au$), $a_{in}\approx 0.6$, which is well below our sampled range of 1--10~au. This explains why, for massive close-in planets, $F_\mathrm{Protection}$ shows little sensitivity to the exact value of $a_\mathrm{ast}$. For lower-mass or more distant planets, however, $a_{in}$ shifts to larger values. In these regimes, our fixed 1--10au range serves as a conservative benchmark. While these planets are less effective at shielding the inner system, Eq.\eqref{equa:safro} implies they would still efficiently eject bodies originating from more distant reservoirs (e.g., a Kuiper-belt-like analog). \textcolor{black}{By maintaining a consistent [1, 10]~au range as justified in Section~\ref{subsec:setup}, we ensure a uniform comparison across the planetary parameter space, while acknowledging that the absolute protection rate for the smallest planets may increase if more distant polluters are considered.}

\subsection{Comparison with previous works\label{subsec:compare}}

\textcolor{black}{Our finding that close-in giant planets serve as a ``pollution filter" is consistent with recent dynamical studies of specific white dwarf systems. For instance, \citet{rogers_wd0141-675_2023} investigated the WD 0141-675 system (prior to the retraction of its planetary candidate), finding that for asteroids on high-eccentricity orbits ($e=0.75$), a massive planet ($9.26 M_\mathrm{Jup}$) at 0.172 au could eject approximately 52\% of them while intercepting 13\% via collisions. Similarly, our simulations for a $13.8~M_\mathrm{Jup}$ planet at 0.1~au show that a total of 77.4\% of asteroids are removed via ejection, with a planetary collision fraction of 3.7\%. The differences in these branching ratios stem from two primary methodological factors. First, the higher collision rates reported by \citet{rogers_wd0141-675_2023} likely result from their specific setup of injecting particles directly into the planetary chaotic zone. Second, while they adopted the Roche radius as the collision criterion, our model allows asteroids to survive multiple Roche-limit passages, significantly increasing their probability of being eventually ejected by the planet. Despite these technical differences, both studies underscore that close-in massive planets act as efficient barriers, significantly suppressing the pollution rate by either ejecting asteroids or intercepting them before they can reach the white dwarf. }

Consequently, our results suggest that this planetary protection might be even more potent than previously estimated for other reservoirs. For example, \citet{pham_polluting_2024} also investigated the influence of a WD1856+534~b like planet on the pollution rate from exo‑Oort‑cloud comets. They reported that a Jupiter-mass planet at 0.02au reduces the accretion rate by at most 50\%, a limit primarily derived from the planet's geometric capacity to intercept comets via physical collisions. In contrast, our simulations show that a similar planet can remove roughly 90\% of small bodies, indicating that the dynamical "scattering barrier" is more efficient than simple physical interception. This discrepancy arises because the planet's influence extends well beyond its physical cross-section, especially for bodies that survive their initial Roche-limit passage. As summarized in Table~\ref{tab:population} for a $1~M_\mathrm{Jup}$ planet at 0.02au, approximately 70\% of the total population is eventually scattered out of the system only after penetrating the Roche limit. Our analysis of individual trajectories (e.g., Fig.\ref{fig:orbits}(b),(c)) confirms that these highly eccentric bodies often remain trapped in the planet's gravitational domain post-passage. Even in cases of tidal disruption, the resulting fragments are more likely to be ejected than accreted. Therefore, we conclude that the protective role of close-in planets is dominated by this sustained dynamical clearing rather than mere physical collision, making the planetary filter significantly more robust than estimates based solely on physical interception.

\section{Conclusion\label{sec:conclu}}

In this work we used N-body simulations to quantify the “pollution shielding” provided by a planet orbiting a white dwarf. By statistically tracking the outcomes of high-eccentricity asteroids in the inner system, we evaluated how a companion modifies their routes to the white dwarf and thereby reduces contamination. Our main conclusions are:

\begin{itemize}
  \item \textbf{Close-in Planets can reduce the fraction of asteroids reaching the Roche limit.} For a close-in companion with mass $13.8\ M_{\rm Jup}$ at $a_p = 0.02\ \mathrm{au}$, over 10\% asteroids are barricaded out of the white dwarf's Roche radius compared to the no-planet case; the planet alters asteroid orbital evolution and lowers the incidence of direct tidal disruption.
  \item \textbf{Most bodies that enter the Roche region are subsequently removed from the system.} Even the asteroids are scattered into the Roche limit or tidally disrupted, the resulting fragments still have a high probability of being dynamically ejected rather than accreted onto the white dwarf.
  \item \textbf{A WD1856+534b-like planet is capable of reducing pollution to one-tenth of its original level.} Our simulations indicate that a close-in planet with properties similar to WD1856+534b can lower the expected contamination rate to one-tenth of its original level. The scattering shielding is even more effective when the incoming pollutant population is not coplanar with the planet.
  \item \textbf{The effectiveness of planetary protection decreases with increasing planet semi-major axis and decreasing planet mass.} At $a_p \simeq 0.02\ \mathrm{au}$, a planet of only $\sim 1\ M_{\rm Jup}$ can eject $\sim 90\%$ of incoming asteroids. At larger separations, substantially higher masses are needed to achieve comparable protection.
\end{itemize}

We conclude that planets and companions can play an important role in regulating white-dwarf pollution by both preventing tidal encounters and removing disrupted fragments. The absence of detectable pollution on WD1856+534 is consistent with predictions of planetary protection. However, with the current limited sample we still cannot statistically distinguish whether this is caused by planetary protection or is merely an observational coincidence. Future systematic discoveries of more similar systems will enable quantitative tests of these theoretical expectations.

\begin{acknowledgements} We would like to thank the referee for their constructive comments, and Siyi Xu for helpful feedback that improved this manuscript. We thank KeTing Shin and Huigen Liu for their valuable discussions on methods, analyses, and figures.
This work is supported by the National Key R\&D Program of China (Grant No. 2024YFA1611803) and the National Natural Science Foundation of China (Grant Nos. 12273011, 12150009 and 12403071). J.-W.X. acknowledges support from the National Youth Talent Support Program. Funding for LAMOST (www.lamost.org) has been provided by the Chinese NDRC. LAMOST is operated and managed by the National Astronomical Observatories, CAS.

\end{acknowledgements}

%
\bibliographystyle{aa} 
\bibliography{Reference} 

@misc{rogers_wd0141-675_2023,
	title = {{WD0141}-675: {A} case study on how to follow-up astrometric planet candidates around white dwarfs},
	shorttitle = {{WD0141}-675},
	url = {http://arxiv.org/abs/2310.05778},
	doi = {10.48550/arXiv.2310.05778},
	language = {en-US},
	urldate = {2025-11-17},
	publisher = {arXiv},
	author = {Rogers, Laura K. and Debes, John and Anslow, Richard J. and Bonsor, Amy and Casewell, S. L. and Santos, Leonardo A. Dos and Dufour, Patrick and Gänsicke, Boris and Fusillo, Nicola Gentile and Koester, Detlev and Nielsen, Louise Dyregaard and Penoyre, Zephyr and Rickman, Emily L. and Sahlmann, Johannes and Tremblay, Pier-Emmanuel and Vanderburg, Andrew and Xu, Siyi and Dennihy, Erik and Farihi, Jay and Hermes, J. J. and Hodgkin, Simon and Kilic, Mukremin and Kowalski, Piotr M. and Sanderson, Hannah and Toonen, Silvia},
	month = oct,
	year = {2023},
	note = {arXiv:2310.05778 [astro-ph]},
}

@article{li_accretion_2021,
	title = {Accretion of tidally disrupted asteroids onto white dwarfs: direct accretion versus disk processing},
	volume = {508},
	issn = {0035-8711, 1365-2966},
	shorttitle = {Accretion of tidally disrupted asteroids onto white dwarfs},
	url = {http://arxiv.org/abs/2106.00441},
	doi = {10.1093/mnras/stab2949},
	language = {en-US},
	number = {4},
	urldate = {2025-02-19},
	journal = {Monthly Notices of the Royal Astronomical Society},
	author = {Li, Daohai and Mustill, Alexander J. and Davies, Melvyn B.},
	month = oct,
	year = {2021},
	note = {arXiv:2106.00441 [astro-ph]},
	pages = {5671--5686},
}

@article{muller_mass-radius_2024,
	title = {The mass-radius relation of exoplanets revisited {\textbar} {Astronomy} \& {Astrophysics} ({A}\&{A})},
	volume = {686},
	copyright = {© The Authors 2024},
	issn = {0004-6361},
	url = {https://www.aanda.org/articles/aa/full_html/2024/06/aa48690-23/aa48690-23.html},
	language = {en-gb},
	urldate = {2025-11-26},
	journal = {Astronomy \& Astrophysics},
	author = {Müller, Simon and Baron, Jana and Helled, Ravit and Bouchy, François and Parc, Léna},
	pages = {A296},
     year = {2024},
}

@article{li_can_2025,
	title = {Can tidal evolution lead to close-in planetary bodies around white dwarfs - {I}. {Orbital} period distribution},
	volume = {537},
	issn = {0035-8711},
	url = {https://ui.adsabs.harvard.edu/abs/2025MNRAS.537.2214L/abstract},
	doi = {10.1093/mnras/staf182},
	language = {en},
	number = {2},
	urldate = {2025-02-16},
	journal = {Monthly Notices of the Royal Astronomical Society, Volume 537, Issue 2, pp.2214-2231},
	author = {Li, Yuqi and Bonsor, Amy and Shorttle, Oliver and Rogers, Laura K.},
	month = feb,
	year = {2025},
	pages = {2214},
}

@article{casewell_phl_2024,
	title = {{PHL} {5038AB}: is the brown dwarf causing pollution of its white dwarf host star?},
	volume = {530},
	issn = {0035-8711},
	shorttitle = {{PHL} {5038AB}},
	url = {https://ui.adsabs.harvard.edu/abs/2024MNRAS.530.3302C},
	doi = {10.1093/mnras/stae974},
	language = {en-US},
	urldate = {2025-10-10},
	journal = {Monthly Notices of the Royal Astronomical Society},
	author = {Casewell, S. L. and Debes, J. and Dupuy, T. J. and Dufour, P. and Bonsor, A. and Rebassa-Mansergas, A. and Murillo-Ojeda, R. and French, J. R. and Alexander, R. D. and Xu, Siyi and Martin, E. and Manjavacas, E.},
	month = may,
	year = {2024},
	note = {Publisher: OUP
ADS Bibcode: 2024MNRAS.530.3302C},
	pages = {3302--3309},
}

@article{blackman_jovian_2021,
	title = {A {Jovian} analogue orbiting a white dwarf star},
	volume = {598},
	copyright = {2021 The Author(s), under exclusive licence to Springer Nature Limited},
	issn = {1476-4687},
	url = {https://www.nature.com/articles/s41586-021-03869-6},
	doi = {10.1038/s41586-021-03869-6},
	language = {en},
	number = {7880},
	urldate = {2025-01-09},
	journal = {Nature},
	author = {Blackman, J. W. and Beaulieu, J. P. and Bennett, D. P. and Danielski, C. and Alard, C. and Cole, A. A. and Vandorou, A. and Ranc, C. and Terry, S. K. and Bhattacharya, A. and Bond, I. and Bachelet, E. and Veras, D. and Koshimoto, N. and Batista, V. and Marquette, J. B.},
	month = oct,
	year = {2021},
	note = {Number: 7880
Publisher: Nature Publishing Group},
	pages = {272--275},
}

@misc{debes_metal_2025,
	title = {Metal {Polluted} {White} {Dwarfs} with 21 $\mu$m {IR} excesses from {JWST}/{MIRI}: {Planets} or {Dust}?},
	shorttitle = {Metal {Polluted} {White} {Dwarfs} with 21 $\mu$m {IR} excesses from {JWST}/{MIRI}},
	url = {http://arxiv.org/abs/2506.21224},
	doi = {10.48550/arXiv.2506.21224},
	urldate = {2025-07-20},
	publisher = {arXiv},
	author = {Debes, John H. and Poulsen, Sabrina and Messier, Ashley and Mullally, Susan E. and Thibault, Katherine and Albert, Loïc and Cracraft, Misty and Bourdais, Érika Le and Dufour, Patrick and Barclay, Tom and Hermes, J. J. and Kilic, Mukremin and Lafrenière, David and Mullally, Fergal and Reach, William and Quintana, Elisa},
	month = jun,
	year = {2025},
	note = {arXiv:2506.21224 [astro-ph]},
}

@misc{pham_polluting_2024,
	title = {Polluting {White} {Dwarfs} with {Oort} {Cloud} {Comets}},
	url = {http://arxiv.org/abs/2404.07160},
	doi = {10.1093/mnras/stae986},
	urldate = {2024-04-12},
	author = {Pham, Dang and Rein, Hanno},
	month = apr,
	year = {2024},
	note = {arXiv:2404.07160 [astro-ph]},
}

@article{chen_power-law_2019,
	title = {A power-law decay evolution scenario for polluted single white dwarfs},
	volume = {3},
	copyright = {2018 The Author(s), under exclusive licence to Springer Nature Limited},
	issn = {2397-3366},
	url = {https://www.nature.com/articles/s41550-018-0609-7},
	doi = {10.1038/s41550-018-0609-7},
	language = {en},
	number = {1},
	urldate = {2025-05-07},
	journal = {Nature Astronomy},
	author = {Chen, Di-Chang and Zhou, Ji-Lin and Xie, Ji-Wei and Yang, Ming and Zhang, Hui and Liu, Hui-Gen and Liang, En-Si and Yu, Zhou-Yi and Yang, Jia-Yi},
	month = jan,
	year = {2019},
	note = {Publisher: Nature Publishing Group},
	pages = {69--75},
}

@article{rein_ias15_2015,
	title = {ias15: a fast, adaptive, high-order integrator for gravitational dynamics, accurate to machine precision over a billion orbits},
	volume = {446},
	issn = {0035-8711},
	shorttitle = {ias15},
	url = {https://doi.org/10.1093/mnras/stu2164},
	doi = {10.1093/mnras/stu2164},
	number = {2},
	urldate = {2024-11-18},
	journal = {Monthly Notices of the Royal Astronomical Society},
	author = {Rein, Hanno and Spiegel, David S.},
	month = jan,
	year = {2015},
	pages = {1424--1437},
}

@article{alonso_transmission_2021,
	title = {A transmission spectrum of the planet candidate {WD} 1856+534 b and a lower limit to its mass},
	volume = {649},
	copyright = {https://www.edpsciences.org/en/authors/copyright-and-licensing},
	issn = {0004-6361, 1432-0746},
	url = {https://www.aanda.org/10.1051/0004-6361/202140359},
	doi = {10.1051/0004-6361/202140359},
	urldate = {2024-12-20},
	journal = {Astronomy \& Astrophysics},
	author = {Alonso, R. and Rodríguez-Gil, P. and Izquierdo, P. and Deeg, H. J. and Lodieu, N. and Cabrera-Lavers, A. and Hollands, M. A. and Pérez-Toledo, F. M. and Castro-Rodríguez, N. and Reverte Payá, D.},
	month = may,
	year = {2021},
	pages = {A131},
}

@article{rein_rebound_2012,
	title = {{REBOUND}: an open-source multi-purpose {N}-body code for collisional dynamics},
	volume = {537},
	copyright = {© ESO, 2012},
	issn = {0004-6361, 1432-0746},
	shorttitle = {{REBOUND}},
	url = {https://www.aanda.org/articles/aa/abs/2012/01/aa18085-11/aa18085-11.html},
	doi = {10.1051/0004-6361/201118085},
	language = {en},
	urldate = {2024-11-27},
	journal = {Astronomy \& Astrophysics},
	author = {Rein, H. and Liu, S.-F.},
	month = jan,
	year = {2012},
	note = {Publisher: EDP Sciences},
	pages = {A128},
}

@article{limbach_miri_2024,
	title = {The {MIRI} {Exoplanets} {Orbiting} {White} dwarfs ({MEOW}) {Survey}: {Mid}-infrared {Excess} {Reveals} a {Giant} {Planet} {Candidate} around a {Nearby} {White} {Dwarf}},
	volume = {973},
	issn = {2041-8205},
	shorttitle = {The {MIRI} {Exoplanets} {Orbiting} {White} dwarfs ({MEOW}) {Survey}},
	url = {https://dx.doi.org/10.3847/2041-8213/ad74ed},
	doi = {10.3847/2041-8213/ad74ed},
	language = {en},
	number = {1},
	urldate = {2024-11-22},
	journal = {The Astrophysical Journal Letters},
	author = {Limbach, Mary Anne and Vanderburg, Andrew and Venner, Alexander and Blouin, Simon and Stevenson, Kevin B. and MacDonald, Ryan J. and Jenkins, Sydney and Bowens-Rubin, Rachel and Soares-Furtado, Melinda and Morley, Caroline and Janson, Markus and Debes, John and Xu, Siyi and Kleisioti, Evangelia and Kenworthy, Matthew and Butler, Paul and Crane, Jeffrey D. and Osip, Dave and Shectman, Stephen and Teske, Johanna},
	month = sep,
	year = {2024},
	note = {Publisher: The American Astronomical Society},
	pages = {L11},
}

@article{carry_density_2012,
	series = {Solar {System} science before and after {Gaia}},
	title = {Density of asteroids},
	volume = {73},
	issn = {0032-0633},
	url = {https://www.sciencedirect.com/science/article/pii/S0032063312000773},
	doi = {10.1016/j.pss.2012.03.009},
	number = {1},
	urldate = {2024-11-21},
	journal = {Planetary and Space Science},
	author = {Carry, B.},
	month = dec,
	year = {2012},
	pages = {98--118},
}

@article{mullally_jwst_2024,
	title = {{JWST} {Directly} {Images} {Giant} {Planet} {Candidates} {Around} {Two} {Metal}-polluted {White} {Dwarf} {Stars}},
	volume = {962},
	issn = {2041-8205},
	url = {https://dx.doi.org/10.3847/2041-8213/ad2348},
	doi = {10.3847/2041-8213/ad2348},
	language = {en},
	number = {2},
	urldate = {2024-11-18},
	journal = {The Astrophysical Journal Letters},
	author = {Mullally, Susan E. and Debes, John and Cracraft, Misty and Mullally, Fergal and Poulsen, Sabrina and Albert, Loic and Thibault, Katherine and Reach, William T. and Hermes, J. J. and Barclay, Thomas and Kilic, Mukremin and Quintana, Elisa V.},
	month = feb,
	year = {2024},
	note = {Publisher: The American Astronomical Society},
	pages = {L32},
}

@article{gansicke_accretion_2019,
	title = {Accretion of a giant planet onto a white dwarf star},
	volume = {576},
	copyright = {2019 The Author(s), under exclusive licence to Springer Nature Limited},
	issn = {1476-4687},
	url = {https://www.nature.com/articles/s41586-019-1789-8},
	doi = {10.1038/s41586-019-1789-8},
	language = {en},
	number = {7785},
	urldate = {2024-11-18},
	journal = {Nature},
	author = {Gänsicke, Boris T. and Schreiber, Matthias R. and Toloza, Odette and Fusillo, Nicola P. Gentile and Koester, Detlev and Manser, Christopher J.},
	month = dec,
	year = {2019},
	note = {Publisher: Nature Publishing Group},
	pages = {61--64},
}

@article{luhman_discovery_2011,
	title = {{DISCOVERY} {OF} {A} {CANDIDATE} {FOR} {THE} {COOLEST} {KNOWN} {BROWN} {DWARF}*},
	volume = {730},
	issn = {2041-8205},
	url = {https://dx.doi.org/10.1088/2041-8205/730/1/L9},
	doi = {10.1088/2041-8205/730/1/L9},
	language = {en},
	number = {1},
	urldate = {2024-11-18},
	journal = {The Astrophysical Journal Letters},
	author = {Luhman, K. L. and Burgasser, A. J. and Bochanski, J. J.},
	month = feb,
	year = {2011},
	note = {Publisher: The American Astronomical Society},
	pages = {L9},
}

@article{thorsett_psr_1993,
	title = {{PSR} {B1620}-26: {A} {Binary} {Radio} {Pulsar} with a {Planetary} {Companion}?},
	volume = {412},
	issn = {0004-637X},
	shorttitle = {{PSR} {B1620}-26},
	url = {https://ui.adsabs.harvard.edu/abs/1993ApJ...412L..33T},
	doi = {10.1086/186933},
	urldate = {2024-11-18},
	journal = {The Astrophysical Journal},
	author = {Thorsett, S. E. and Arzoumanian, Z. and Taylor, J. H.},
	month = jul,
	year = {1993},
	note = {Publisher: IOP
ADS Bibcode: 1993ApJ...412L..33T},
	pages = {L33},
}

@article{sigurdsson_young_2003,
	title = {A {Young} {White} {Dwarf} {Companion} to {Pulsar} {B1620}-26: {Evidence} for {Early} {Planet} {Formation}},
	volume = {301},
	shorttitle = {A {Young} {White} {Dwarf} {Companion} to {Pulsar} {B1620}-26},
	url = {https://www.science.org/doi/10.1126/science.1086326},
	doi = {10.1126/science.1086326},
	number = {5630},
	urldate = {2024-11-18},
	journal = {Science},
	author = {Sigurdsson, Steinn and Richer, Harvey B. and Hansen, Brad M. and Stairs, Ingrid H. and Thorsett, Stephen E.},
	month = jul,
	year = {2003},
	note = {Publisher: American Association for the Advancement of Science},
	pages = {193--196},
}

@article{jura_carbon_2006,
	title = {Carbon {Deficiency} in {Externally} {Polluted} {White} {Dwarfs}: {Evidence} for {Accretion} of {Asteroids}},
	volume = {653},
	issn = {0004-637X},
	shorttitle = {Carbon {Deficiency} in {Externally} {Polluted} {White} {Dwarfs}},
	url = {https://iopscience.iop.org/article/10.1086/508738/meta},
	doi = {10.1086/508738},
	language = {en},
	number = {1},
	urldate = {2024-11-18},
	journal = {The Astrophysical Journal},
	author = {Jura, M.},
	month = dec,
	year = {2006},
	note = {Publisher: IOP Publishing},
	pages = {613},
}

@article{zuckerman_aluminumcalcium-rich_2011,
	title = {{AN} {ALUMINUM}/{CALCIUM}-{RICH}, {IRON}-{POOR}, {WHITE} {DWARF} {STAR}: {EVIDENCE} {FOR} {AN} {EXTRASOLAR} {PLANETARY} {LITHOSPHERE}?},
	volume = {739},
	issn = {0004-637X},
	shorttitle = {{AN} {ALUMINUM}/{CALCIUM}-{RICH}, {IRON}-{POOR}, {WHITE} {DWARF} {STAR}},
	url = {https://dx.doi.org/10.1088/0004-637X/739/2/101},
	doi = {10.1088/0004-637X/739/2/101},
	language = {en},
	number = {2},
	urldate = {2024-11-18},
	journal = {The Astrophysical Journal},
	author = {Zuckerman, B. and Koester, D. and Dufour, P. and Melis, Carl and Klein, B. and Jura, M.},
	month = sep,
	year = {2011},
	note = {Publisher: The American Astronomical Society},
	pages = {101},
}

@article{zuckerman_chemical_2007,
	title = {The {Chemical} {Composition} of an {Extrasolar} {Minor} {Planet}},
	volume = {671},
	issn = {0004-637X},
	url = {https://iopscience.iop.org/article/10.1086/522223/meta},
	doi = {10.1086/522223},
	language = {en},
	number = {1},
	urldate = {2024-11-18},
	journal = {The Astrophysical Journal},
	author = {Zuckerman, B. and Koester, D. and Melis, C. and Hansen, Brad M. and Jura, M.},
	month = dec,
	year = {2007},
	note = {Publisher: IOP Publishing},
	pages = {872},
}

@article{jura_x-ray_2009,
	title = {X-{RAY} {AND} {INFRARED} {OBSERVATIONS} {OF} {TWO} {EXTERNALLY} {POLLUTED} {WHITE} {DWARFS}},
	volume = {699},
	issn = {0004-637X},
	url = {https://dx.doi.org/10.1088/0004-637X/699/2/1473},
	doi = {10.1088/0004-637X/699/2/1473},
	language = {en},
	number = {2},
	urldate = {2024-11-18},
	journal = {The Astrophysical Journal},
	author = {Jura, M. and Muno, M. P. and Farihi, J. and Zuckerman, B.},
	month = jun,
	year = {2009},
	note = {Publisher: The American Astronomical Society},
	pages = {1473},
}

@article{jura_tidally_2003,
	title = {A {Tidally} {Disrupted} {Asteroid} around the {White} {Dwarf} {G29}-38},
	volume = {584},
	issn = {0004-637X},
	url = {https://iopscience.iop.org/article/10.1086/374036/meta},
	doi = {10.1086/374036},
	language = {en},
	number = {2},
	urldate = {2024-11-18},
	journal = {The Astrophysical Journal},
	author = {Jura, M.},
	month = jan,
	year = {2003},
	note = {Publisher: IOP Publishing},
	pages = {L91},
}

@article{debes_are_2002,
	title = {Are {There} {Unstable} {Planetary} {Systems} around {White} {Dwarfs}?},
	volume = {572},
	issn = {0004-637X},
	url = {https://iopscience.iop.org/article/10.1086/340291/meta},
	doi = {10.1086/340291},
	language = {en},
	number = {1},
	urldate = {2024-11-18},
	journal = {The Astrophysical Journal},
	author = {Debes, John H. and Sigurdsson, Steinn},
	month = jun,
	year = {2002},
	note = {Publisher: IOP Publishing},
	pages = {556},
}

@misc{zuckerman_ancient_2010,
	title = {Ancient planetary systems are orbiting a large fraction of white dwarf stars},
	url = {http://arxiv.org/abs/1007.2252},
	urldate = {2024-11-17},
	publisher = {arXiv},
	author = {Zuckerman, B. and Melis, C. and Klein, B. and Koester, D. and Jura, M.},
	month = jul,
	year = {2010},
	note = {arXiv:1007.2252},
}

@misc{zhu_exoplanet_2021,
	title = {Exoplanet {Statistics} and {Theoretical} {Implications}},
	url = {http://arxiv.org/abs/2103.02127},
	doi = {10.48550/arXiv.2103.02127},
	urldate = {2024-11-17},
	publisher = {arXiv},
	author = {Zhu, Wei and Dong, Subo},
	month = mar,
	year = {2021},
	note = {arXiv:2103.02127},
}

@article{coutu_analysis_2019,
	title = {Analysis of {Helium}-rich {White} {Dwarfs} {Polluted} by {Heavy} {Elements} in the {Gaia} {Era}},
	volume = {885},
	issn = {0004-637X},
	url = {https://dx.doi.org/10.3847/1538-4357/ab46b9},
	doi = {10.3847/1538-4357/ab46b9},
	language = {en},
	number = {1},
	urldate = {2024-11-17},
	journal = {The Astrophysical Journal},
	author = {Coutu, S. and Dufour, P. and Bergeron, P. and Blouin, S. and Loranger, E. and Allard, N. F. and Dunlap, B. H.},
	month = nov,
	year = {2019},
	note = {Publisher: The American Astronomical Society},
	pages = {74},
}

@article{veras_smallest_2023,
	title = {The smallest planetary drivers of white dwarf pollution},
	volume = {519},
	issn = {0035-8711, 1365-2966},
	url = {http://arxiv.org/abs/2301.04160},
	doi = {10.1093/mnras/stad130},
	number = {4},
	urldate = {2024-03-25},
	journal = {Monthly Notices of the Royal Astronomical Society},
	author = {Veras, Dimitri and Rosengren, Aaron J.},
	month = jan,
	year = {2023},
	note = {arXiv:2301.04160 [astro-ph]},
	pages = {6257--6266},
}

@article{veras_orbit_2022,
	title = {Orbit decay of 2-100 au planetary remnants around white dwarfs with no gravitational assistance from planets},
	volume = {510},
	issn = {0035-8711},
	url = {https://ui.adsabs.harvard.edu/abs/2022MNRAS.510.3379V},
	doi = {10.1093/mnras/stab3490},
	urldate = {2025-02-23},
	journal = {Monthly Notices of the Royal Astronomical Society},
	author = {Veras, Dimitri and Birader, Yusuf and Zaman, Uwais},
	month = mar,
	year = {2022},
	note = {Publisher: OUP
ADS Bibcode: 2022MNRAS.510.3379V},
	pages = {3379--3388},
}

@article{vanderburg_giant_2020,
	title = {A giant planet candidate transiting a white dwarf},
	volume = {585},
	copyright = {2020 The Author(s), under exclusive licence to Springer Nature Limited},
	issn = {1476-4687},
	url = {https://www.nature.com/articles/s41586-020-2713-y},
	doi = {10.1038/s41586-020-2713-y},
	language = {en},
	number = {7825},
	urldate = {2023-12-28},
	journal = {Nature},
	author = {Vanderburg, Andrew and Rappaport, Saul A. and Xu, Siyi and Crossfield, Ian J. M. and Becker, Juliette C. and Gary, Bruce and Murgas, Felipe and Blouin, Simon and Kaye, Thomas G. and Palle, Enric and Melis, Carl and Morris, Brett M. and Kreidberg, Laura and Gorjian, Varoujan and Morley, Caroline V. and Mann, Andrew W. and Parviainen, Hannu and Pearce, Logan A. and Newton, Elisabeth R. and Carrillo, Andreia and Zuckerman, Ben and Nelson, Lorne and Zeimann, Greg and Brown, Warren R. and Tronsgaard, René and Klein, Beth and Ricker, George R. and Vanderspek, Roland K. and Latham, David W. and Seager, Sara and Winn, Joshua N. and Jenkins, Jon M. and Adams, Fred C. and Benneke, Björn and Berardo, David and Buchhave, Lars A. and Caldwell, Douglas A. and Christiansen, Jessie L. and Collins, Karen A. and Colón, Knicole D. and Daylan, Tansu and Doty, John and Doyle, Alexandra E. and Dragomir, Diana and Dressing, Courtney and Dufour, Patrick and Fukui, Akihiko and Glidden, Ana and Guerrero, Natalia M. and Guo, Xueying and Heng, Kevin and Henriksen, Andreea I. and Huang, Chelsea X. and Kaltenegger, Lisa and Kane, Stephen R. and Lewis, John A. and Lissauer, Jack J. and Morales, Farisa and Narita, Norio and Pepper, Joshua and Rose, Mark E. and Smith, Jeffrey C. and Stassun, Keivan G. and Yu, Liang},
	month = sep,
	year = {2020},
	note = {Number: 7825
Publisher: Nature Publishing Group},
	pages = {363--367},
}

@article{frewen_eccentric_2014,
	title = {Eccentric planets and stellar evolution as a cause of polluted white dwarfs},
	volume = {439},
	issn = {0035-8711},
	url = {https://doi.org/10.1093/mnras/stu097},
	doi = {10.1093/mnras/stu097},
	number = {3},
	urldate = {2024-11-18},
	journal = {Monthly Notices of the Royal Astronomical Society},
	author = {Frewen, S. F. N. and Hansen, B. M. S.},
	month = apr,
	year = {2014},
	pages = {2442--2458},
}

@misc{koester_frequency_2014,
	title = {The frequency of planetary debris around young white dwarfs},
	url = {http://arxiv.org/abs/1404.2617},
	doi = {10.48550/arXiv.1404.2617},
	urldate = {2024-11-17},
	publisher = {arXiv},
	author = {Koester, Detlev and Gänsicke, Boris T. and Farihi, Jay},
	month = apr,
	year = {2014},
	note = {arXiv:1404.2617},
}

@misc{veras_evolution_2024,
	title = {The evolution and delivery of rocky extra-solar materials to white dwarfs},
	url = {http://arxiv.org/abs/2401.08767},
	doi = {10.48550/arXiv.2401.08767},
	urldate = {2024-02-23},
	publisher = {arXiv},
	author = {Veras, Dimitri and Mustill, Alexander J. and Bonsor, Amy},
	month = jan,
	year = {2024},
	note = {arXiv:2401.08767 [astro-ph, physics:physics]},
}

@article{xu__chemical_2017,
	title = {The {Chemical} {Composition} of an {Extrasolar} {Kuiper}-{Belt}-{Object}},
	volume = {836},
	issn = {2041-8213},
	url = {https://iopscience.iop.org/article/10.3847/2041-8213/836/1/L7},
	doi = {10.3847/2041-8213/836/1/L7},
	language = {en},
	number = {1},
	urldate = {2023-10-25},
	journal = {The Astrophysical Journal},
	author = {Xu , S. and Zuckerman, B. and Dufour, P. and Young, E. D. and Klein, B. and Jura, M.},
	month = feb,
	year = {2017},
	pages = {L7},
}

@book{veras_planetary_2021,
  author    = {Veras, Dimitri},
  title     = {Planetary Systems Around White Dwarfs},
  publisher = {Oxford University Press},
  year      = {2021},
  note      = {In Oxford Research Encyclopedia of Planetary Science},
  doi       = {10.1093/acrefore/9780190647926.013.238},
  url       = {https://oxfordre.com/planetaryscience/view/10.1093/acrefore/9780190647926.001.0001/acrefore-9780190647926-e-238},
  urldate   = {2023-10-27}
}

@article{oconnor_pollution_2023,
	title = {On the pollution of white dwarfs by exo-{Oort} cloud comets},
	volume = {524},
	issn = {0035-8711, 1365-2966},
	url = {http://arxiv.org/abs/2306.10102},
	doi = {10.1093/mnras/stad2281},
	number = {4},
	urldate = {2023-10-15},
	journal = {Monthly Notices of the Royal Astronomical Society},
	author = {O'Connor, Christopher E. and Lai, Dong and Seligman, Darryl Z.},
	month = jul,
	year = {2023},
	note = {arXiv:2306.10102 [astro-ph]},
	pages = {6181--6197},
}

@article{debes_link_2012,
	title = {{THE} {LINK} {BETWEEN} {PLANETARY} {SYSTEMS}, {DUSTY} {WHITE} {DWARFS}, {AND} {METAL}-{POLLUTED} {WHITE} {DWARFS}},
	volume = {747},
	issn = {0004-637X},
	url = {https://dx.doi.org/10.1088/0004-637X/747/2/148},
	doi = {10.1088/0004-637X/747/2/148},
	language = {en},
	number = {2},
	urldate = {2024-11-18},
	journal = {The Astrophysical Journal},
	author = {Debes, John H. and Walsh, Kevin J. and Stark, Christopher},
	month = feb,
	year = {2012},
	note = {Publisher: The American Astronomical Society},
	pages = {148},
}

@article{zuckerman_metal_2003,
	title = {Metal {Lines} in {DA} {White} {Dwarfs}*},
	volume = {596},
	issn = {0004-637X},
	url = {https://iopscience.iop.org/article/10.1086/377492/meta},
	doi = {10.1086/377492},
	language = {en},
	number = {1},
	urldate = {2024-11-17},
	journal = {The Astrophysical Journal},
	author = {Zuckerman, B. and Koester, D. and Reid, I. N. and Hünsch, M.},
	month = oct,
	year = {2003},
	note = {Publisher: IOP Publishing},
	pages = {477},
}

@article{mcdonald_binary_2023,
	title = {Binary asteroid scattering around white dwarfs},
	volume = {520},
	issn = {0035-8711, 1365-2966},
	url = {http://arxiv.org/abs/2302.00020},
	doi = {10.1093/mnras/stad382},
	number = {3},
	urldate = {2023-11-07},
	journal = {Monthly Notices of the Royal Astronomical Society},
	author = {McDonald, Catriona H. and Veras, Dimitri},
	month = feb,
	year = {2023},
	note = {arXiv:2302.00020 [astro-ph]},
	pages = {4009--4022},
}

@misc{limbach_thermal_2025,
	title = {Thermal {Emission} and {Confirmation} of the {Frigid} {White} {Dwarf} {Exoplanet} {WD} 1856+534b},
	url = {http://arxiv.org/abs/2504.16982},
	doi = {10.48550/arXiv.2504.16982},
	urldate = {2025-04-27},
	publisher = {arXiv},
	author = {Limbach, Mary Anne and Vanderburg, Andrew and MacDonald, Ryan J. and Stevenson, Kevin B. and Jenkins, Sydney and Blouin, Simon and Rauscher, Emily and Bowens-Rubin, Rachel and Gallo, Elena and Mang, James and Morley, Caroline V. and Sing, David K. and O'Connor, Christopher and Venner, Alexander and Xu, Siyi},
	month = apr,
	year = {2025},
	note = {arXiv:2504.16982 [astro-ph]},
}

@article{xu_geminigmos_2021,
	title = {Gemini/{GMOS} {Transmission} {Spectroscopy} of the {Grazing} {Planet} {Candidate} {WD} 1856+534 b},
	volume = {162},
	issn = {1538-3881},
	url = {https://dx.doi.org/10.3847/1538-3881/ac2d26},
	doi = {10.3847/1538-3881/ac2d26},
	language = {en},
	number = {6},
	urldate = {2025-05-09},
	journal = {The Astronomical Journal},
	author = {Xu, Siyi and Diamond-Lowe, Hannah and MacDonald, Ryan J. and Vanderburg, Andrew and Blouin, Simon and Dufour, P. and Gao, Peter and Kreidberg, Laura and Leggett, S. K. and Mann, Andrew W. and Morley, Caroline V. and Stephens, Andrew W. and O’Connor, Christopher E. and Thao, Pa Chia and Lewis, Nikole K.},
	month = dec,
	year = {2021},
	note = {Publisher: The American Astronomical Society},
	pages = {296},
}

@article{veras_white_2020,
	title = {The white dwarf planet {WD} {J0914}+1914 b: barricading potential rocky pollutants?},
	volume = {493},
	copyright = {https://academic.oup.com/journals/pages/open\_access/funder\_policies/chorus/standard\_publication\_model},
	issn = {0035-8711, 1365-2966},
	shorttitle = {The white dwarf planet {WD} {J0914}+1914 b},
	url = {https://academic.oup.com/mnras/article/493/4/4692/5804793},
	doi = {10.1093/mnras/staa625},
	language = {en},
	number = {4},
	urldate = {2025-05-22},
	journal = {Monthly Notices of the Royal Astronomical Society},
	author = {Veras, Dimitri},
	month = apr,
	year = {2020},
	pages = {4692--4699},
}

@article{paquette_diffusion_1986,
	title = {Diffusion {Coefficients} for {Stellar} {Plasmas}},
	volume = {61},
	issn = {0067-0049},
	url = {https://ui.adsabs.harvard.edu/abs/1986ApJS...61..177P},
	doi = {10.1086/191111},
	urldate = {2025-05-29},
	journal = {The Astrophysical Journal Supplement Series},
	author = {Paquette, C. and Pelletier, C. and Fontaine, G. and Michaud, G.},
	month = may,
	year = {1986},
	note = {Publisher: IOP
ADS Bibcode: 1986ApJS...61..177P},
	pages = {177},
}

@article{swan_interpretation_2019,
	title = {Interpretation and diversity of exoplanetary material orbiting white dwarfs},
	volume = {490},
	issn = {0035-8711, 1365-2966},
	url = {http://arxiv.org/abs/1908.08047},
	doi = {10.1093/mnras/stz2337},
	number = {1},
	urldate = {2025-05-31},
	journal = {Monthly Notices of the Royal Astronomical Society},
	author = {Swan, Andrew and Farihi, Jay and Koester, Detlev and Hollands, Mark and Parsons, Steven and Cauley, P. Wilson and Redfield, Seth and Gaensicke, Boris T.},
	month = nov,
	year = {2019},
	note = {arXiv:1908.08047 [astro-ph]},
	pages = {202--218},
}

@article{swan_planetesimals_2023,
	title = {Planetesimals at {DZ} stars – {I}. {Chondritic} compositions and a massive accretion event},
	volume = {526},
	issn = {0035-8711},
	url = {https://doi.org/10.1093/mnras/stad2867},
	doi = {10.1093/mnras/stad2867},
	number = {3},
	urldate = {2025-05-31},
	journal = {Monthly Notices of the Royal Astronomical Society},
	author = {Swan, Andrew and Farihi, Jay and Melis, Carl and Dufour, Patrick and Desch, Steven J and Koester, Detlev and Guo, Jincheng},
	month = dec,
	year = {2023},
	pages = {3815--3831},
}

@article{malamud_tidal_2020,
	title = {Tidal disruption of planetary bodies by white dwarfs {I}: a hybrid sph-analytical approach},
	volume = {492},
	issn = {0035-8711},
	shorttitle = {Tidal disruption of planetary bodies by white dwarfs {I}},
	url = {https://doi.org/10.1093/mnras/staa142},
	doi = {10.1093/mnras/staa142},
	number = {4},
	urldate = {2024-12-05},
	journal = {Monthly Notices of the Royal Astronomical Society},
	author = {Malamud, Uri and Perets, Hagai B},
	month = mar,
	year = {2020},
	pages = {5561--5581},
}
%
\end{document}